\documentclass[journal, comsoc]{IEEEtran}

\usepackage{algorithm}
\usepackage{algorithmic}
\usepackage{amsmath}
\usepackage{amsfonts}
\usepackage{amssymb}
\usepackage{array}
\usepackage{bbding}
\usepackage{booktabs}
\usepackage{cite}
\usepackage{color}
\usepackage{diagbox}
\usepackage{enumerate}
\usepackage{graphicx}
\usepackage{multirow}
\usepackage{setspace}
\usepackage{textcomp}
\usepackage{multirow}
\usepackage{url}
\usepackage{threeparttable, tablefootnote}
\usepackage{enumerate}
\ifCLASSOPTIONcompsoc \usepackage[caption=false, font=normalsize, labelfont=sf, textfont=sf]{subfig}
\else
\usepackage[caption=false, font=footnotesize]{subfig}
\fi
\usepackage[numbers,sort&compress]{natbib}

\allowdisplaybreaks[1]

\newtheorem{theorem}{Theorem}

\newcommand{\tabincell}[2]{\begin{tabular}{@{}#1@{}}#2\end{tabular}}

\begin{document}

\title{Service Chain Composition with Failures in NFV Systems: A Game-Theoretic Perspective}

\author{Simeng~Bian,
    Xi~Huang,
    Ziyu~Shao*,
    Xin~Gao,
    Yang~Yang  
    
\thanks{S. Bian, X. Huang, Z. Shao, X. Gao, Y. Yang are with the School of Information Science and Technology, ShanghaiTech University, Shanghai 201210, China. (E-mail: {biansm, huangxi, shaozy, gaoxin, yangyang}@shanghaitech.edu.cn) (*Corresponding author: Ziyu Shao)}
}

\maketitle

\begin{abstract}
For state-of-the-art network function virtualization (NFV) systems, it remains a key challenge to conduct effective \textit{service chain composition} for different network services (NSs) with ultra-low request latencies and minimum network congestion.
To this end, existing solutions often require full knowledge of the network state, while ignoring the privacy issues and overlooking the non-cooperative behaviors of users. What is more, they may fall short in the face of unexpected failures such as user unavailability and virtual machine breakdown.
In this paper, we formulate the problem of service chain composition in NFV systems with failures as a \textit{non-cooperative game}. 
By showing that such a game is a \textit{weighted potential game} and exploiting the unique problem structure, we propose two effective distributed schemes that guide the service chain compositions of different NSs towards the Nash equilibrium (NE) state with both near-optimal latencies and minimum congestion.
Besides, we develop two novel learning-aided schemes as comparisons, which are based on deep reinforcement learning (DRL) and Monte Carlo tree search (MCTS) techniques, respectively.
Our theoretical analysis and simulation results demonstrate the effectiveness of our proposed schemes, as well as the adaptivity when faced with failures. 
\end{abstract}

\begin{IEEEkeywords}
	NFV, service chain composition, quality of service, non-cooperative game, deep reinforcement learning, Monte Carlo tree search.
\end{IEEEkeywords}

\IEEEpeerreviewmaketitle

\section{Introduction}
\label{sec:introduction}
\IEEEPARstart{I}{n} the face of ever-increasing and diversified demands of user requests for network services (NSs), network function virtualization (NFV) techniques are the key to enable flexible and efficient service delivery \cite{mijumbi2015network}.
Instead of deploying each NS on dedicated hardwares, NFV systems virtualize each type of network function as multiple instances on different commodity servers, then chain the instances of different virtual network functions (VNFs) in a specified order at runtime to carry out particular NS, \textit{a.k.a.} \textit{service chain composition} \cite{herrera2016resource}. 

\begin{table*}[!t]
	\small
	\centering
	\caption{Comparisons between existing solutions and our proposed schemes}
	\label{tab: comparison}
    \begin{tabular}{|l|c|c|c|c|c|c|c|c|c|c|}
    \hline
    ~ & \multicolumn{2}{c|}{Optimization Based} & \multicolumn{4}{c|}{Game Theory Based} & \multicolumn{4}{c|}{Learning Based} \\
	\cline{2-11}
    ~ & ~~~\cite{bari2015orchestrating}~~~ 
    & ~~~\cite{wang2015dynamic}~~~ 
    & \cite{d2017exploiting} 
    & \cite{le2020congestion} 
    & MA-SCCA 
    & MH-SCCA 
    & \cite{xiao2019nfv}
    & \cite{soualah2017link}
    & DRL-SCCA 
    & MCTS-SCCA 
    \\
    \hline  
    Performance Guarantee & - & $\checkmark$ & $\checkmark$ & $\checkmark$ & $\checkmark$ & $\checkmark$ & - & - & - & -\\
    \hline
    Distributed Implementation & - & $\checkmark$ & $\checkmark$ & - & $\checkmark$ & $\checkmark$ & - & -  & - & - \\
    \hline
	{Non-cooperative Users} 
    & - & - & $\checkmark$ & - & $\checkmark$ & $\checkmark$ & - & - & $\checkmark$ & $\checkmark$\\
    \hline
    Failure Awareness & - & - & - & - & $\checkmark$ & $\checkmark$ & - & - & $\checkmark$ & $\checkmark$\\
    \hline
    Decision-making Overheads$^*$ & M & M & H & H & H & L & M & L & M & L\\
    \hline
    \end{tabular}
    \begin{tablenotes}
	\footnotesize
	\item[1] $^*$ H, M, and L represent high, medium, and low decision-making overheads, respectively.
    \end{tablenotes}
    \vspace{-.3cm}
\end{table*}

\begin{figure}[!t]
    \centering
    \includegraphics[scale=.28]{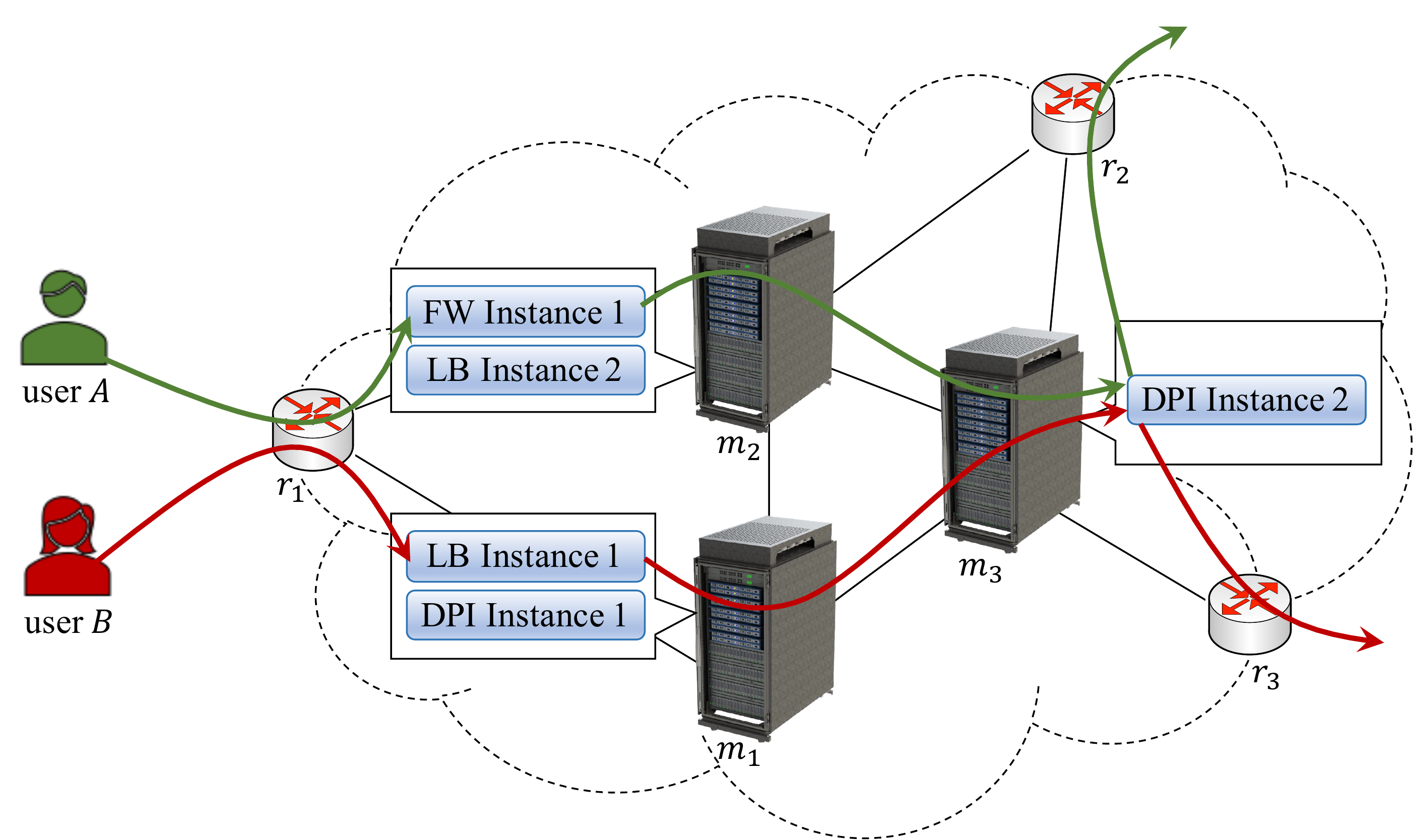}
    \caption{An example of service chain composition.}
    \label{fig: model}
\end{figure}

We show an example in Figure \ref{fig: model} to illustrate the service chain composition in detail. 
The NFV system provides three types of VNFs: firewall (FW), deep packet inspection (DPI), and load balancer (LB). Their instances are distributed on three servers $\{m_1, m_2, m_3\}$. 
Meanwhile, there are two users communicating with the system through gateway routers $\{r_1, r_2, r_3\}$. User $A$ subscribes the NS of service chain FW $\to$ DPI, while the service chain of user $B$ is LB $\to$ DPI.  
Taking user $B$ as an example, we see that it has four possible service chain composition choices, because there are two LB instances and two DPI instances in the system. In Figure \ref{fig: model}, it chooses LB Instance 1 and DPI Instance 2. The red arrows represent the corresponding path of its traffic. 
Different composition choices may incur different request latencies and different levels of network congestion. Intuitively, for user $B$, choosing LB Instance 1 and DPI Instance 1 may be a better choice. The reasons include 1) compared with data transmission between two servers ($m_1$ and $m_3$), data transmission within a single server ($m_1$) will result in a \textit{shorter latency}; and 2) compared with resource contention on DPI Instance 2, monopolizing DPI Instance 1 will result in a \textit{lower congestion}. 

To date, it still remains an $\mathcal{NP}$-hard problem to conduct effective service chain composition in NFV systems with various NSs to achieve ultra-low request latencies and minimum network congestion \cite{cohen2015near}. 
Moreover, in practice, connections from users to NFV systems may become unexpectedly unavailable \cite{gill2011understanding}, and VNF instances may also crash at runtime \cite{birke2014failure}. Such issues make the problem even more complicated.  
In this paper, we consider the problem of service chain composition in NFV systems with both user and VNF instance failures from a game-theoretic perspective.
We summarize our key results and contributions as follows.

\textbf{Modeling and Formulation:} 
We formulate the service chain composition problem with both user and VNF instance failures as a \textit{non-cooperative game} among users, aiming to reduce request latencies and mitigate network congestion. 
Further more, we show that the formulated game is a \textit{weighted potential game}; \textit{i.e.}, any unilateral local improvement made by an individual user will lead to a global improvement in the overall system performance. 
Such a property makes it attainable to achieve the optimal service chain composition solution in a fully distributed manner. 
To the best of knowledge, this paper is the first to study the service chain composition problem with both user and VNF instance failures from the game-theoretic perspective. 
    
\textbf{Scheme Design and Performance Analysis:} 
With such a formulation, we adopt Markov approximation techniques \cite{chen2013markov} and propose an effective \textit{distributed} scheme called MA-SCCA (Markov Approximation based Service Chain Composition Algorithm). 
MA-SCCA achieves provably near-optimal request latencies and network congestion through interactions among users. 
To further accelerate the decision making process, we adopt Metropolis-Hastings methods \cite{chib1995understanding} and propose MH-SCCA (Metropolis-Hastings based Service Chain Composition Algorithm), a variant of MA-SCCA which achieves comparable performance with a lower complexity.

\textbf{Learning-aided Schemes:} 
As comparisons to MA-SCCA and MH-SCCA, we propose two learning-aided service chain composition schemes: DRL-SCCA (Deep Reinforcement Learning based Service Chain Composition Algorithm) and MCTS-SCCA (Monte Carlo Tree Search based Service Chain Composition Algorithm). 
To our best knowledge, we are the first to adopt deep reinforcement learning (DRL) \cite{li2017deep} and Monte Carlo tree search (MCTS) \cite{browne2012survey} techniques to search for the optimal Nash equilibrium, respectively.

\textbf{Experimental Verification and Analysis:}
We conduct extensive simulations to evaluate the performance of our proposed schemes. 
The results validate their effectiveness, as well as the adaptivity of MA-SCCA and MH-SCCA when faced with user and VNF instance failures. 

The rest of the paper is organized as follows. 
We discuss related works in Section \ref{sec: related work}.
Then, in Section \ref{sec: problem formulation} we present our model and problem formulation, followed by the proof of its game property.  
Section \ref{sec: scheme design} elaborates the design of MA-SCCA and MH-SCCA with their corresponding performance analysis. 
Next, Section \ref{sec: learning-aided algo design} shows the detailed designs of DRL-SCCA and MCTS-SCCA.
We evaluate the performance of the proposed schemes based on numerical simulations in Section \ref{sec: numerical results} and conclude this paper in Section \ref{sec: conclusion}. 

\section{Related Works}  
\label{sec: related work}
The service chain composition problem is a critical but challenging resource management problem in the field of NFV. 
By far, a great number of works have been conducted \cite{mijumbi2015network}. We compare our proposed schemes with some of the state-of-the-art solutions in Table \ref{tab: comparison}.

\subsection{Optimization Based Approaches}
Most of existing works solve the service chain composition problem from the optimization perspective, focusing on system energy cost minimization  \cite{bari2015orchestrating, beck2017scalable} or network utility maximization  \cite{wang2015dynamic} for the Telecom Operator. Due to the $\mathcal{NP}$-hardness of the problem \cite{mijumbi2015network}, these works often resort to approximation or heuristic methods to reduce the complexity of the proposed algorithms. For example, Bari \textit{et al}. \cite{bari2015orchestrating} formulated the problem as an integer linear programming (ILP) problem, then developed a heuristic method based on Viterbi algorithm. Wang \textit{et al}.  \cite{wang2015dynamic} equated the problem as a multi-dimension multi-choice knapsack problem, and adopted Markov approximation \cite{chen2013markov} techniques to devise a distributed algorithm. Beck \textit{et al}. \cite{beck2017scalable} proposed a heuristic algorithm that coordinates the composition of service chains and their embeddings into the substrate network.

Though the effectiveness of such works has been well justified, they do not consider the following aspects.  
\begin{itemize}
    \item They ignore the individualistic behaviors of users.  Specifically, they assume that users are \textit{cooperative} in utility optimization for the Telecom Operator. However, in practice, users are usually \textit{non-cooperative}, aiming to optimize their own benefits. 
    \item They require a rich set of prior information including user locations, service chain requests, and traffic rates. However, considering \textit{privacy issues}, some users may be unwilling to share such information with others.
\end{itemize}
With all these concerns in mind, in this paper, we model the service chain composition problem from a game-theoretic perspective, and devise distributed schemes that jointly consider the non-cooperative behaviors and privacy of users.

\subsection{Game Theory Based Approaches}
To the best of our knowledge, D'Oro \textit{et al.} \cite{d2017exploiting} are the first to study the service chain composition problem from the game-theoretic perspective. They modeled the problem as a non-cooperative \textit{congestion game} and developed a distributed algorithm which provably converges to a Nash equilibrium (NE) through interactions among users. Such a design preserves the privacy of users.
Following the work in \cite{d2017exploiting}, Le \textit{et al.} \cite{le2020congestion} proposed a centralized algorithm that achieves the best NE with the optimal system performance. However, their solution is essentially based on the brute-force search, with a high computational complexity. Chiang \textit{et al.} \cite{chiang2019distributed} modeled the problem as an \textit{exact potential game} by regrading each server as a player. The game eventually ends up with an NE, since every finite exact potential game possesses an NE \cite{monderer1996potential}.

Inspired by all these works, this paper also takes a game-theoretic perspective. However, our work differs from theirs in the following aspects.
First, we take the user and VNF instance failures into account. By exploiting the unique problem structure, we formulate the problem as a non-cooperative \textit{potential game}, with the aim to find the \textit{optimal} NE. Our proposed schemes, namely MA-SCCA and MH-SCCA, are both distributed and privacy-preserving. Moreover, they can adapt to failures in a timely manner. 

\subsection{Learning Based Approaches}
In addition, some works leveraged recently prevailing learning techniques such as deep reinforcement learning (DRL) \cite{li2017deep} and Monte Carlo tree search (MCTS) \cite{browne2012survey} to solve the service chain composition problem. 
For instance, Xiao \textit{et al.} \cite{xiao2019nfv} proposed NFVDeep, a DRL-based approach that deploys service chains with different QoS requirements and adopts serialization-and-backtracking methods to shrink the action space size. Khezri \textit{et al.}  \cite{khezri2019deep} designed a DQN-based agent that arranges service chains to maximize the admission ratio and minimize the placement cost.
In the meanwhile, Soualah \textit{et al.} \cite{soualah2017link} adopted Monte Carlo tree search methods and proposed a link failure recovery algorithm. Their solution focuses on selecting and re-mapping reliable paths for network services so as to minimize the negative effects of link failures. 

Inspired by all these works, in this paper we propose two centralized schemes, namely DRL-SCCA and MCTS-SCCA, that solve the service chain composition problem based on DRL and MCTS methods, respectively. Unlike the above works, our proposed schemes focus on searching for the best Nash equilibrium state.
Under such a state, the system performance is optimal, \textit{i.e.}, with ultra-low request latencies and minimum network congestion. 

\section{Problem Formulation}  
\label{sec: problem formulation}
In this section, we first introduce our system model. Then we present the game-theoretic formulation of the problem. By exploiting the problem structure, we further analyze the property of the game. 

\subsection{System Model}
We consider an NFV system owned by a Telecom Operator that provides network services to users. Table \ref{tab: notations} summarizes the key notations in this paper, including the values of notations in Figure \ref{fig: model}.
                                                                                                        
\textbf{Deployment Model:}
The system provides a set $\mathcal{F}$ of available types of VNFs, with $F \triangleq |\mathcal{F}|$. 
Inside the system, there is a server cluster $\mathcal{M}$, with $M \triangleq |\mathcal{M}|$. 
Each server $m \in \mathcal{M}$ holds a set of VNF instances $\mathcal{F}_m$.\footnote{We assume that a server has at most one instance for each VNF type. However, our model can be directly extended to the case where a server has multiple instances for some types of VNFs.} Each VNF instance is deployed on a standalone virtual machine (VM).\footnote{In the rest of this paper, we use the terms \textit{VNF instance} and \textit{VM} interchangeably.} 
We use $\mathcal{V} \triangleq $ $ \bigcup_{m \in \mathcal{M}} \mathcal{F}_{m}$ to denote the set of all VNF instances (VMs) in the system. 
Besides servers, the system also includes a set of gateway routers $\mathcal{R}$ which manage the communication with external networks. 
For any two nodes $u_1, u_2 \in \mathcal{R} \cup \mathcal{V}$, we use $l_{\left(u_1, u_2\right)}$ to denote the latency between them. 

\textbf{Users and Network Services:}
Outside the system, there is a group of users, denoted by a set $\mathcal{N}$. 
The total number of users is $N \triangleq |\mathcal{N}|$. 
Each user $i \in \mathcal{N}$ subscribes a specific network service (NS) $\mathcal{F}_i$, denoted by an ordered VNF chain $\{f_{i,1}, f_{i,2}, \cdots, f_{i,F_{i}}\}$ of length $F_i \! \triangleq \! |\mathcal{F}_{i}|$. At runtime, user $i$ generates traffic with a rate of $\lambda_i$. The traffic first flows into the system through an ingress router $r^{\text{(in)}}_i \! \in \! \mathcal{R}$, then flows through VNF instances of $f_{i, j}$ for $j \! = \! \{1, 2, \cdots, F_i\}$, and finally leaves the system from an egress router $r^{\text{(out)}}_i \! \in \! \mathcal{R}$. 

\textbf{Service Chain Composition Decisions:}
We denote $\mathbf{w}_i$ as the service chain composition decision made by user $i$. The equality $\mathrm{w}_{i,j} = v$ indicates that user $i$ chooses VM $v$ to execute its $j$-th VNF.  
By denoting $\mathbf{w}_{-i}$ as the decisions of all users other than $i$, we can write $\mathbf{w} \triangleq (\mathbf{w}_{i}, \mathbf{w}_{-i})$ as the combination of all users' decisions.

\textbf{Failures:}
In practice, users may be unavailable due to various reasons \cite{gill2011understanding}. Likewise, VMs may break down unexpectedly at runtime \cite{birke2014failure}. 
In this paper, we consider the simplest case in which such failures are independently and identically distributed.\footnote{Our model can serve as a basis to handle more general scenarios with non-uniform and dependent failures.} Specifically, we assume that all users have identical survival and failure probabilities of $\gamma_{\text{(user)}}$ and $\bar{\gamma}_{\text{(user)}}$, respectively; and that all VMs have identical survival and failure probabilities of $\gamma_{\text{(VM)}}$ and $\bar{\gamma}_{\text{(VM)}}$, respectively.

\begin{table}[!t]
	\caption{Key Notations}
	\label{tab: notations}
	\small
	\centering
    \begin{tabular}{|c|l|}
    \hline
    \textbf{Notations} & \textbf{Descriptions and Examples} \\
    \hline
    $\mathcal{F}$ & Set of VNF types, $\mathcal{F} = \{\text{FW}, \text{LB}, \text{DPI}\}$ \\
    \hline
    $F$ & Number of VNF types, $F = 3$ \\
    \hline
    $\mathcal{M}$ & Set of servers, $\mathcal{M} = \{m_1, m_2, m_3\}$ \\
    \hline
    $M$ & Number of servers, $M = 3$ \\
    \hline
    $\mathcal{F}_m$ & \tabincell{l}{Set of VNF instances provided by server $m$ \\ $\mathcal{F}_{m_1} = \{\text{LB Instance 1}, \text{DPI Instance 1}\}$} \\
    \hline
    $\mathcal{V}$ & \tabincell{l}{Set of all VNF instances (VMs) \\ $\mathcal{V} = \{\text{FW Instance 1}, \cdots, \text{DPI Instance 2}\}$} \\
    \hline
    $\mathcal{R}$ & Set of gateway routers, $\mathcal{R} = \{r_1, r_2, r_3\}$ \\
    \hline
    $l_{\left(u_1, u_2\right)}$ & \tabincell{l}{Latency between node $u_1$ and $u_2$ \\ $l_{(r_1, \text{FW Instance 1})} = 1ms$}\\
    \hline
    $\mathcal{N}$ & Set of users, $\mathcal{N} = \{A, B\}$\\
    \hline
    $N$ & Number of users, $N = 2$ \\
    \hline
    $\mathcal{F}_i$ & \tabincell{l}{Network service request of user $i$ \\ $\mathcal{F}_{A} = \{\text{FW}, \text{DPI}\}$}\\
    \hline
    $f_{i,j}$ & \tabincell{l}{$j$-th VNF in user $i$'s required service chain \\ $f_{A, 1} = \text{FW}, f_{A, 2} = \text{DPI}$} \\
    \hline
    $F_i$ & Length of user $i$'s service chain, $F_A = 2$\\
    \hline
    $\lambda_i$ & Traffic rate of user $i$, $\lambda_A = 10Mbps$ \\
    \hline
    $r^{\text{(in)}}_i$ & Ingress router of user $i$, $r^{\text{(in)}}_A = r_1$ \\
    \hline
    $r^{\text{(out)}}_i$ & Egress router of user $i$, $r^{\text{(out)}}_A = r_2$ \\
    \hline
    $\gamma_{\text{(user)}}$ & Survival probability of users, $\gamma_{\text{(user)}} = 0.9$ \\
    \hline
    $\bar{\gamma}_{\text{(user)}}$ & Failure probability of users, $\bar{\gamma}_{\text{(user)}} = 0.1$ \\
    \hline
	$\gamma_{\text{(VM)}}$ & Survival probability of VMs, $\gamma_{\text{(VM)}} = 0.9$ \\
    \hline
    $\bar{\gamma}_{\text{(VM)}}$ & Failure probability of VMs, $\bar{\gamma}_{\text{(VM)}} = 0.1$ \\
    \hline
    $\Omega_i$ & Failure cost of user $i$, $\Omega_A = 1000$ units\\
    \hline
    $\mathbf{w}_i$ & \tabincell{l}{Service chain composition decision of user $i$ \\ $\mathbf{w}_A = \{\text{FW Instance 1}, \text{DPI Instance 2}\}$} \\
    \hline
    $\mathrm{w}_{i, j}$ & \tabincell{l}{The VM chosen for the $j$-th VNF of user $i$ \\ $\mathrm{w}_{A, 1} = \text{FW Instance 1}$} \\
    \hline
    \end{tabular}
\end{table} 

\subsection{User Costs}
Different service chain composition decisions may result in different costs. Formally, we define the cost function of user $i$ as follows. 

\begin{enumerate}[i)]
	\item If user $i$ fails, it will receive a large cost (penalty) $\Omega_i$. Such a case happens with probability $\bar{\gamma}_{\text{(user)}}$. 
	\item If user $i$ survives, but at least one of its chosen VMs fails, the corresponding cost is also $\Omega_i$. Such a case occurs with probability $\gamma_{\text{(user)}} \cdot \left(1 - \gamma_{\text{(VM)}}^{F_i}\right)$.
	\item If both user $i$ and all its selected VMs survive, which occurs with probability $\gamma_{\text{(user)}} \cdot \gamma_{\text{(VM)}}^{F_i}$, we should consider both latency and congestion costs.
	\begin{itemize}
	    \item \textbf{Total Latency:} Given the decision $\mathbf{w}_{i}$, we define the total latency experienced by user $i$ as
		\begin{equation}\label{eq: def_latency}
    		c_{i}^{(\text{L})}(\mathbf{w}_i) \triangleq l_{\left(r^{\text{(in)}}_i, \mathrm{w}_{i,1}\right)} + \sum_{j=2}^{F_i} l_{\left(\mathrm{w}_{i,j-1}, \mathrm{w}_{i,j}\right)} + l_{\left(\mathrm{w}_{i, F_i}, r^{\text{(out)}}_i\right)}.
		\end{equation}
		
		\item \textbf{Total Congestion:} Different users may employ the same VM. Therefore, we should also take the congestion due to such resource contention into account. Given the decisions of all users $\mathbf{w}$, we define the set of users that share VM $v$ as 
		\begin{equation}\label{eq: def_M}
    		\mathcal{N}_{v}(\mathbf{w}) \triangleq \{k | k\in \mathcal{N}, v \in \mathbf{w}_k \}.
		\end{equation}
		Next, given that user $i$ survives, we define the workload on VM $v \in \mathbf{w}_i$ as
		\begin{equation}\label{eq: def_delta}
    		\delta_{i, v}(\mathbf{w}) \triangleq \lambda_i ~+ \sum_{k\in \mathcal{N}_{v}(\mathbf{w}) / \{i\}} \gamma_{\text{(user)}} \cdot \lambda_k. 
		\end{equation} 
		Then the total congestion experienced by user $i$ is
		\begin{equation}\label{eq: def_congestion}
    		c_{i}^{(\text{C})}(\mathbf{w}) \triangleq \sum_{v \in \mathbf{w}_i} \delta_{i, v}(\mathbf{w}).
		\end{equation}
	\end{itemize}
	Therefore, if both user $i$ and all its selected VMs survive, the cost experienced by user $i$ is $\alpha c_i^{(\text{L})}(\mathbf{w}_i) + c_i^{(\text{C})}(\mathbf{w})$. The parameter $\alpha$ is a positive constant that weighs the relative importance of latency reduction to congestion mitigation.
\end{enumerate}

Combining the costs defined in i), ii), and iii), the expected cost of user $i$ is given by
\begin{equation}\label{eq: cost_function}
\begin{split}
    c_i(\mathbf{w}) \triangleq & ~\bar{\gamma}_{\text{(user)}} \cdot {\Omega_i} + \gamma_{\text{(user)}} \cdot \left(1- \gamma_{\text{(VM)}}^{F_i} \right) \cdot \Omega_i \\
    & + \gamma_{\text{(user)}} \cdot \gamma_{\text{(VM)}}^{F_i} \cdot  \left[ \alpha c_i^{(\text{L})}(\mathbf{w}_i) + c_i^{(\text{C})}(\mathbf{w})  \right].
\end{split}
\end{equation}

\subsection{Game-Theoretic Formulation}
In this section, we formulate the service chain composition problem as a \textit{non-cooperative} game. Typically, a game consists of three elements: a set of players, a set of strategies, and cost (utility) functions. 
\begin{itemize}
    \item \textbf{Player}: We view each user of the NFV system as a player, and denote $\mathcal{N}$ as the set of players. 
    \item \textbf{Strategy}: Each player's strategy set is denoted by $\mathcal{W}_i$, which includes all its possible service chain composition decisions. We use $\mathcal{W} \triangleq \bigcup_{i \in \mathcal{N}}\mathcal{W}_i$ to denote the joint strategy space of all players.
    \item \textbf{Cost Function}: The cost function $c_i(\cdot)$ for each player $i$ is defined in (\ref{eq: cost_function}).
\end{itemize}

We then formulate a game $\mathcal{G}$ which is characterized by the triad $(\mathcal{N}, \mathcal{W}, \{c_i\}_{i\in \mathcal{N}})$.
In such a game, each player $i$ aims to minimize its own cost. Such a cost is not only determined by its own strategy $\mathbf{w}_i$ but also other players' strategies $\mathbf{w}_{-i}$. 
The goal of the game is to search for a Nash equilibrium (NE), a state in which no player has the incentive to change its current strategy unilaterally. 
However, when the number of players is huge, it is in general challenging to find an NE. 
Furthermore, there may be more than one equilibrium in the game and each equilibrium may induce different system performances. 
For the Telecom Operator, it is often preferable to achieve the optimal system performance, which adds even more complexities to the design of the service chain composition schemes.
In the following, we exploit the unique structure of the problem and show that the two goals of finding an NE and optimizing the system performance can actually be unified. 

We first show that the game $\mathcal{G}$ is a \textit{weighted potential game}. In such a game, the local performance improvement in terms of cost reduction by any individual player $i$ is \textit{proportional} to the global performance improvement of the system.
Specifically, we have the following theorem.
\begin{theorem}
	\textit{
    The game $\mathcal{G}$ is a weighted potential game with potential function
    \begin{equation}\label{eq: potential}
    \begin{split}
        \Phi(\mathbf{w}) &\triangleq 2\alpha \gamma_{\text{{\upshape (user)}}} \sum_{k\in \mathcal{N}} \lambda_k c_k^{(\text{L})}(\mathbf{w}_k) + \sum_{v \in \mathcal{V}} \Big(\sum_{k \in \mathcal{N}_v(\mathbf{w})} \lambda_k \gamma_{\text{{\upshape (user)}}} \Big)^2,\\
    \end{split}
    \end{equation}
    such that for any pair of states $\mathbf{w} = (\mathbf{w}_i, \mathbf{w}_{-i})$ and $\mathbf{w}' = (\mathbf{w}'_i, \mathbf{w}_{-i})$ that involves only player $i$'s strategy change, 
    \begin{equation}\label{eq: potential_cost}
        \Phi(\mathbf{w}) - \Phi(\mathbf{w}') = 2\lambda_i \gamma_{\text{{\upshape (VM)}}}^{- F_i} \left( {c}_i(\mathbf{w}) - {c}_i(\mathbf{w}') \right).
    \end{equation}
    }
\label{theorem: potential}
\end{theorem}

The proof is relegated to Appendix-A.
For the potential function defined in (\ref{eq: potential}), its first term is the summation of the product of each player's traffic rate and the latency it experiences. Such a term actually represents the overall traffic load in the system. The second term is the square-sum of the expected workloads of all VMs. Such a term reflects the congestion status of the system. Therefore, we can view the weighted sum of the overall traffic load and congestion load, \textit{i.e.} the potential function $\Phi(\cdot)$, as one measurement of the overall system performance. The Telecom Operator aims to optimize the system performance, \textit{i.e.}, minimize the potential value. Each individual player aims to improve its own benefits, \textit{i.e.}, minimize its cost. In the following, we show that in such a weighted potential game, the two goals actually coincide. 

Formally, the Telecom Operator tries to solve the following optimization problem:
\begin{equation}\label{eq: problem_initial}
\begin{split}
    \underset{\mathbf{w}}{\text{Minimize}} ~~~~ &\Phi(\mathbf{w})\\
    \text{Subject to} ~~~~ &\mathbf{w}  \in \mathcal{W}.
\end{split}
\end{equation}
We denote $\mathbf{w}^*$ as the optimal solution to problem (\ref{eq: problem_initial}). Then, we show that $\mathbf{w}^*$ is also an NE of game $\mathcal{G}$. Let's consider the scenario where the current service chain composition decision is $\mathbf{w}^* = (\mathbf{w}^*_1, \cdots, \mathbf{w}^*_N)$ and player $i$ plans to change its strategy from $\mathbf{w}^*_i$ to $\mathbf{w}_i$. Since $\mathbf{w}^*$ is the optimal solution to problem (\ref{eq: problem_initial}), we have
\begin{equation} \label{eq: phi_diff}
    \Phi(\mathbf{w}_i, \mathbf{w}^*_{-i}) - \Phi(\mathbf{w}^*_i, \mathbf{w}^*_{-i}) \geq 0.
\end{equation}
Combining (\ref{eq: phi_diff}) and Theorem \ref{theorem: potential}, we obtain
\begin{equation} \label{eq: cost_diff}
    c_i(\mathbf{w}_i, \mathbf{w}^*_{-i}) -c_i(\mathbf{w}^*_i, \mathbf{w}^*_{-i}) \geq 0.
\end{equation}
Inequality (\ref{eq: cost_diff}) holds for any $\mathbf{w}_i \in \mathcal{W}_i$, indicating that player $i$ can not attain any benefits (cost reduction) by changing its strategy from $\mathbf{w}_i^*$ to any other. Therefore, $\mathbf{w}^*$ is an NE. 

In the rest of this paper, we refer to the optimal solution $\mathbf{w}^*$ as the optimal NE of game $\mathcal{G}$. Under such a state, the system performance is optimized and no player has the incentive to change its strategy unilaterally, which simultaneously satisfies the goals of the Telecom Operator and all players.

\section{Scheme Design}  
\label{sec: scheme design}
Essentially, problem (\ref{eq: problem_initial}) is a combinatorial problem with its search space size exponential in the number of VNF types and the number of players. Such a problem is in general $\mathcal{NP}$-hard to be solved \cite{cohen2015near}.
For practical NFV systems, it is often desirable to develop easy-to-implement solutions that solve the problem approximately with performance guarantees. 
With such a mentality, we adopt Markov approximation framework \cite{chen2013markov} to solve the service chain composition problem, leading to distributed solutions which approximate the optimal Nash equilibrium (NE) through interactions of players. 
In the following subsections, we first transform problem (\ref{eq: problem_initial}) into an approximation form. Then we develop an effective distributed scheme called MA-SCCA (Markov Approximation based Service Chain Composition Algorithm), as well as its variant called MH-SCCA (Metropolis-Hastings based Service Chain Composition Algorithm). Finally, we conduct performance analysis for the two proposed schemes. 

\subsection{Problem Transformation}
First, we note that problem (\ref{eq: problem_initial}) is actually equivalent to the following problem with the same optimal value: 
\begin{equation}\label{eq: problem_eq}
\begin{split}
    \underset{\mathbf{\pi} \succeq 0}{\text{Minimize}} ~~~~ &\sum_{\mathbf{w} \in \mathcal{W}} \pi_{\mathbf{w}} \Phi(\mathbf{w}) \\
    \text{Subject to} ~~~ &\sum_{\mathbf{w} \in \mathcal{W}} \pi_{\mathbf{w}} = 1,
\end{split}
\end{equation}
where variable $\pi$ is a non-negative vector of length $|\mathcal{W}|$. 
In fact, problem (\ref{eq: problem_eq}) can be viewed as the optimization problem to decide the time fraction $\pi_{\mathbf{w}}$ that the system spends on each state $\mathbf{w}$, so as to minimize the time-average of potential value $\Phi$ over a period of time. 
Therefore, the optimal value of problem (\ref{eq: problem_eq}) is achieved only when the system spends all the time on the state $\mathbf{w}^*$ with minimum potential value, namely the desired optimal state. 

Next, by applying \textit{log-sum-exponential} approximation \cite{chen2013markov}, we approximate problem (\ref{eq: problem_eq}) as follows   
\begin{equation}\label{eq: problem_approx}
\begin{split}
    \underset{\mathbf{\pi} \succeq 0}{\text{Minimize}} ~~~~ &\sum_{\mathbf{w} \in \mathcal{W}} \pi_{\mathbf{w}} \Phi(\mathbf{w}) + \frac{1}{\beta} \sum_{\mathbf{w} \in \mathcal{W}} \pi_{\mathbf{w}} \log \pi_{\mathbf{w}} \\
    \text{Subject to} ~~~ &\sum_{\mathbf{w} \in \mathcal{W}} \pi_{\mathbf{w}} = 1,
\end{split}
\end{equation}
where $\beta$ is a positive constant that regulates the approximation gap, as elaborated by the following theorem.
\begin{theorem}\label{theorem: gap}
	\textit{
    The optimality gap induced by the log-sum-exponential approximation is upper-bounded by $\frac{1}{\beta} FN\log M$.
    }
\end{theorem}

We relegate the proof to Appendix-B. Theorem \ref{theorem: gap} implies that the approximation gap can be reduced by increasing the value of $\beta$. 
By solving the KKT conditions of problem (\ref{eq: problem_approx}), we obtain the optimal solution as
\begin{equation}\label{eq: opt_prob}
    \pi^{*}_{\mathbf{w}} = \frac{\exp [-\beta \Phi(\mathbf{w})]}{\sum_{\bar{\mathbf{w}} \in \mathcal{W}} \exp [-\beta \Phi(\bar{\mathbf{w}})]}, \ \forall \, \mathbf{w} \in \mathcal{W}.
\end{equation}
Then the NFV system can proceed by spending a fraction $\pi^{*}_{\mathbf{w}}$ of time on each state $\mathbf{w}$ to solve problem (\ref{eq: problem_initial}) approximately. 
However, in practice, the size of the search space $\mathcal{W}$ is usually rather huge, making the calculation of (\ref{eq: opt_prob}) computationally intractable.
Instead of direct calculation, we switch to designing a discrete-time Markov chain and adopt random sampling techniques to obtain $\pi^{*}_{\mathbf{w}}$ in an asymptotic fashion \cite{chen2013markov}. 

\subsection{Markov Chain Design}
We construct a discrete-time Markov chain with its state space as $\mathcal{W}$. 
In general, we have two degrees of freedom to design such a Markov chain: its topology and the transition probabilities among states. 
    
\subsubsection{Topology} 
For a Markov chain, its topology is defined as a graph in which each vertex denotes a particular state and each edge implies the mutual accessibility between its two associated states by one-step transition. 
    
For topology of the designed Markov chain, we only consider two types of edges (transitions). 
The first one is \textit{self-loop}, \textit{i.e.}, for each state $\mathbf{w}$, the transition to itself is allowed. 
The other is the transition between any pair of states that differs in exactly one player's strategy change. 
Formally, given any two states $\mathbf{w}$ and $\mathbf{w}'$, they are connected in the topology if and only if their Hamming distance $H_{\mathbf{w}, \mathbf{w}'} \leq 1$. The Hamming distance is defined as follows:
\begin{equation}\label{eq: hamming}
    H_{\mathbf{w}, \mathbf{w}'} \triangleq \sum_{i\in \mathcal{N}} \mathbb{I}_{\mathbf{w}_i \neq \mathbf{w}'_i},
\end{equation}
in which $\mathbb{I}_{A}$ is the indicator function for the occurrence of event $A$. Specifically, $\mathbb{I}_{A} = 1$ if event $A$ occurs and zero otherwise.
 
\subsubsection{Transition Probability} 
On one hand, for each state $\mathbf{w}$, the probability of its self-transition is defined as
\begin{equation}\label{eq: trans_prob_self}
    p_{\mathbf{w}, \mathbf{w}} \triangleq \frac{1}{N} \sum_{i\in \mathcal{N}} \frac{\exp [-\beta \Phi(\mathbf{w})]}{\sum_{\bar{\mathbf{w}} \in \mathcal{A}_i} \exp [-\beta \Phi(\bar{\mathbf{w}})]}.
\end{equation}
Note that $\mathcal{A}_i$ is a subset of the state space that only player $i$'s strategy varies and other players' strategies remain unchanged. Formally, we have $\mathcal{A}_i \triangleq \{(\bar{\mathbf{w}}_i, \mathbf{w}_{-i})~|~ \bar{\mathbf{w}}_i \in \mathcal{W}_i \}$.

On the other hand, for any pair of states $\mathbf{w}$ and $\mathbf{w}'$ that differs in player $i$'s strategy, the transition probability from state $\mathbf{w}$ to state $\mathbf{w}'$ is defined as
\begin{equation}\label{eq: trans_prob_chain}
    p_{\mathbf{w}, \mathbf{w}'} \triangleq \frac{1}{N} \frac{\exp [-\beta \Phi(\mathbf{w}')]}{\sum_{\bar{\mathbf{w}} \in \mathcal{A}_i} \exp [-\beta \Phi(\bar{\mathbf{w}})] }.
\end{equation}
The transition probabilities are zero for other pairs of states.

The designed Markov chain enjoys some properties that facilitate our subsequent scheme designs, as shown by the following theorem.
\begin{theorem}
    \textit{
	    The designed Markov chain is time-reversible with its stationary distribution identical to (\ref{eq: opt_prob}). 
	}
\label{theorem: chain}
\end{theorem}

The proof is relegated to Appendix-C.

\subsection{Scheme Design}
We implement the Markov chain that proceeds over iterations. In each iteration, given current state $\mathbf{w}$, each player $i$ is equally likely to be selected to update its strategy. Once chosen, player $i$ would change its strategy unilaterally from $\mathbf{w}_i$ to $\mathbf{w}'_i$ with probability \footnote{Note that $\mathbf{w}'_i$ may be identical to $\mathbf{w}_i$, which corresponds to a self-transition.} 
\begin{equation}\label{eq: trans_prob_alg}
     p^{(i)}_{(\mathbf{w}_i, \mathbf{w}_{-i}), (\mathbf{w}'_i, \mathbf{w}_{-i})} \triangleq \frac{\exp [-\beta \Phi(\mathbf{w}'_i, \mathbf{w}_{-i})]}{\sum_{\bar{\mathbf{w}}_i \in \mathcal{W}_i} \exp [-\beta \Phi(\bar{\mathbf{w}}_i, \mathbf{w}_{-i})] }.
\end{equation}
For such a general procedure, two questions remain unsolved.

The \textit{first} is regarding the player selection in each iteration. 
Clearly, the selection can be implemented in a centralized manner by employing a dedicated server to maintain all players. Once selected, the only player will get notified and authorized the chance to change its strategy.
Such a centralized implementation requires global information such as the number of players, which may change over time due to network dynamics. Moreover, the dedicated server may be a single point of failure. 
With all these concerns in mind, how can we implement the player selection in a distributed fashion?

The \textit{second} is regarding the state transition in each iteration. 
Note that according to (\ref{eq: potential}), the calculation of the potential value $\Phi$ in (\ref{eq: trans_prob_alg}) requires all players' information, including their strategies and traffic rates.
Of course, we can employ a dedicated server to conduct this calculation. 
However, such an implementation can cause privacy issues, since some players may be unwilling to reveal their personal information.
Moreover, the dedicated server may be a single point of failure.
Then, is it possible to calculate the transition probability (\ref{eq: trans_prob_alg}) by the player itself using only local information?
    
In response to the above two questions, we show the detailed design of player selection and state transition as follows.
    
\textbf{Distributed Player Selection:}
To conduct the player selection procedure in a distributed fashion, we equip each player with a Poisson countdown clock of rate $\mu$. 
At the beginning of each iteration, each player will set up a countdown timer independently with the time length randomly generated from an exponential distribution of mean $1/\mu$. 
The player whose timer first expires would acquire the chance to change its strategy. Meanwhile, it will notify other players to abort their timers. 
For such a procedure, we have the following remarks:
\begin{itemize} \label{remark on scl and adp}
    \item[$\diamond$] The above procedure ensures that, in every iteration, each player has an equal probability to be selected. Furthermore, exactly one of them will be selected.  
    \item[$\diamond$] Besides, such an implementation enjoys a good adaptivity. Specifically, the countdown-clock setup requires no knowledge of the number of players nor information update upon the joining or leaving of players. 
    \item[$\diamond$] In practice, by setting $\mu \gg 1$, the countdown (player selection) procedure will take up only a small time fraction of each iteration. 
\end{itemize}   
    
\textbf{Transition with Local Information:}
    To calculate the transition probability using only local information, we take a transformation as follows. 
    According to (\ref{eq: potential_cost}), for the chosen player $i$, we have
\begin{equation}
\label{eq: trans_prob_alg2}
\begin{split}
    p^{(i)}_{(\mathbf{w}_i, \mathbf{w}_{-i}), (\mathbf{w}'_i, \mathbf{w}_{-i})} &= \frac{\exp [-\beta \Phi(\mathbf{w}'_i, \mathbf{w}_{-i})]}{\sum_{\bar{\mathbf{w}}_i \in \mathcal{W}_i} \exp [-\beta \Phi(\bar{\mathbf{w}}_i, \mathbf{w}_{-i} )] }\\
    &= \frac{1}{\sum_{\bar{\mathbf{w}}_i \in \mathcal{W}_i} \exp \{ \beta [\Phi(\mathbf{w}'_i, \mathbf{w}_{-i}) - \Phi(\bar{\mathbf{w}}_i, \mathbf{w}_{-i})] \} }\\
	&= \frac{\exp [-2\beta  \lambda_i \gamma_{\text{(VM)}}^{-F_i} c_i(\mathbf{w}'_i, \mathbf{w}_{-i})]}{\sum_{\bar{\mathbf{w}}_i \in \mathcal{W}_i} \exp [-2\beta \lambda_i \gamma_{\text{(VM)}}^{-F_i} c_i(\bar{\mathbf{w}}_i, \mathbf{w}_{-i})] }. 
\end{split}
\end{equation}
Unlike (\ref{eq: trans_prob_alg}), the transition probability (\ref{eq: trans_prob_alg2}) depends only on the local information of player $i$ such as its traffic rate $\lambda_i$, its service chain length $F_{i}$, and its cost  $c_{i}$.

\begin{algorithm}[!b]
    \caption{Markov Approximation based Service Chain Composition Algorithm (MA-SCCA) in one iteration for each \textbf{player} \boldmath$i$}
    \begin{algorithmic}[1]
    \label{alg}
    \STATE \textbf{Repeat do}
    \IF {player $i$ receives signal \textit{RESET}}
        \STATE Reset its clock with rate $\mu$.
        \STATE Start the countdown.
    \ENDIF
    \IF {player $i$ receives signal \textit{SUSPEND}}
        \STATE Suspend its clock.
    \ENDIF
    \IF {player $i$'s clock expires}
        \STATE Send signal \textit{SUSPEND} to other players.
        \STATE Apply strategy $\mathbf{w}_i'$ chosen from its strategy set $ \mathcal{W}_i$ with probability $p^{(i)}_{(\mathbf{w}_i, \mathbf{w}_{-i}), (\mathbf{w}'_i, \mathbf{w}_{-i})}$ defined in (\ref{eq: trans_prob_alg2}).
    \ENDIF
    \end{algorithmic}
\end{algorithm}

Based on the above design, we propose \textit{MA-SCCA} (Markov Approximation based Service Chain Composition Algorithm) that solves problem (\ref{eq: problem_approx}), thereby solving problem (\ref{eq: problem_initial}) in an approximate fashion. 
The pseudocode of MA-SCCA is shown in Algorithm \ref{alg}. 
Initially, each player picks a strategy randomly from its strategy space and runs Algorithm \ref{alg}.   
Note that, for line $2$ in Algorithm \ref{alg}, the RESET signal is sent from the Telecom Operator at the beginning of each iteration. 

\textbf{Remarks:} MA-SCCA runs in a fully distributed manner with decision making by individual players with their local information. 
The only necessary message passing in each iteration includes the \textit{RESET} signal from the Telecom Operator and the \textit{SUSPEND} signal from the chosen player. Besides, according to Theorem \ref{theorem: gap}, we can reduce the approximation gap by increasing the value of $\beta$ in (\ref{eq: trans_prob_alg2}). However, this will also lead to a longer convergence time \cite{chen2013markov}. In practice, the parameter $\beta$ should be well-tuned to achieve both good system performance and fast convergence rate.

\subsection{Variant of MA-SCCA}
In each iteration, the chosen player $i$ should calculate the transition probability (\ref{eq: trans_prob_alg2}) for each strategy in its strategy space. To complete this computation, player $i$ should report all its possible strategies $\bar{\mathbf{w}}_i \in \mathcal{W}_i$ to the Telecom Operator, and download the corresponding costs $c_i(\bar{\mathbf{w}}_i, \mathbf{w}_{-i})$ from the Telecom Operator. Such a procedure leads to high data transmission overheads. To mitigate such overheads, we adopt Metropolis-Hastings techniques \cite{chib1995understanding}, and propose a variant of MA-SCCA called \textit{MH-SCCA} (Metropolis-Hastings based Service Chain Composition Algorithm).

The core component of the Metropolis-Hastings design is a Markov chain, which is known as the \textit{proposal} chain. With the proposal chain, the general procedure proceeds as follows.
In each iteration, the proposal chain will propose a state transition from the current state according to its transition probabilities. 
Such a proposal would be accepted with some certain probability, \textit{a.k.a.} \textit{acceptance probability}. If the proposal is accepted, the state transition succeeds. Otherwise, the Markov chain will stay on the original state. 
By a proper design of the transition and acceptance probabilities, we can acquire a Markov chain with the desired stationary distribution (\ref{eq: opt_prob}).
For MH-SCCA, its proposal chain has the same topology as the Markov chain of MA-SCCA. 
In the following, we first show how MH-SCCA proceeds in general and then the design of its transition and acceptance probabilities. 

\textbf{General Procedure of MH-SCCA:}
In each iteration, given the current state $\mathbf{w}$, a player is selected uniformly randomly (following the same way in MA-SCCA) to update its strategy. 
The chosen player $i$ will pick a strategy $\mathbf{w}_i'$ uniformly randomly from its strategy set,\footnote{The picked strategy may be identical to the player's current strategy, which corresponds to a self-transition.} resulting in a state transition proposal from $\mathbf{w} = (\mathbf{w}_i, \mathbf{w}_{-i})$ to $\mathbf{w}' = (\mathbf{w}'_i, \mathbf{w}_{-i})$.
Such a proposal will be accepted with probability $a_{\mathbf{w}, \mathbf{w}'}$, defined as
\begin{equation}\label{eq: accept_prob}
    a_{\mathbf{w}, \mathbf{w}'} = \min\left(\frac{\exp(-\beta \Phi(\mathbf{w}'))}{\exp(-\beta \Phi(\mathbf{w}))}, 1\right).
\end{equation}
Upon acceptance, the player will change its strategy accordingly. Otherwise, it sticks to its previous strategy. 

Different from MA-SCCA, the transition probabilities of the proposal chain of MH-SCCA are as follows.
Given a state $\mathbf{w}$, a self-transition occurs in the proposal chain only when the chosen player picks its original strategy again, with probability 
\begin{equation}\label{eq: trans_prob_self}
    q_{\mathbf{w}, \mathbf{w}} \triangleq \frac{1}{N} \sum_{i\in \mathcal{N}} \frac{1}{|\mathcal{W}_i|}.
\end{equation}
Meanwhile, for any pair of states $\mathbf{w}$ and $\mathbf{w}'$ that differs in only player $i$'s strategy, the transition probability is
\begin{equation}\label{eq: trans_prob_chain}
    q_{\mathbf{w}, \mathbf{w}'} \triangleq \frac{1}{N} \cdot \frac{1}{|\mathcal{W}_i|}.
\end{equation}
The transition probabilities of other state pairs are all zero.
In order to ensure the validity of the proposal chain, we show that, all the transition probabilities out from a given state $\mathbf{w}$ sum to one: 
\begin{equation}
\begin{split}
    \sum_{\mathbf{w}' \in \mathcal{W}} q_{\mathbf{w}, \mathbf{w}'} 
    &= q_{\mathbf{w}, \mathbf{w}} + \sum_{\substack{\mathbf{w}' \in \mathcal{W} \\ H_{\mathbf{w}, \mathbf{w}'}=1}} q_{\mathbf{w}, \mathbf{w}'}\\
    &= \frac{1}{N} \sum_{i\in \mathcal{N}} \frac{1}{|\mathcal{W}_i|} + \sum_{i\in \mathcal{N}} \left(|\mathcal{W}_i| - 1\right) \cdot \frac{1}{N} \cdot \frac{1}{|\mathcal{W}_i|}\\
    &= \sum_{i\in \mathcal{N}} \frac{1}{N} \cdot \frac{1}{|\mathcal{W}_i|} \cdot |\mathcal{W}_i|= 1.
\end{split}
\end{equation}
    
Next, we show that the Markov chain induced by the state proposal and acceptance procedures has the same stationary distribution as (\ref{eq: opt_prob}). 
We prove this by verifying that the detailed balance equations are satisfied for any two states $\mathbf{w}$ and $\mathbf{w}'$ with Hamming distance $H_{\mathbf{w}, \mathbf{w}'}=1$. Specifically,
\begin{equation}
\begin{split}
    &\pi^*_{\mathbf{w}} \cdot q_{\mathbf{w}, \mathbf{w}'} \cdot a_{\mathbf{w}, \mathbf{w}'} \\
    =& \frac{\exp [-\beta \Phi(\mathbf{w})]}{\sum_{\bar{\mathbf{w}} \in \mathcal{W}} \exp [-\beta \Phi(\bar{\mathbf{w}})]} \cdot \frac{1}{N|\mathcal{W}_i|} \cdot \min\left(\frac{\exp(-\beta \Phi(\mathbf{w}'))}{\exp(-\beta \Phi(\mathbf{w}))}, 1\right) \\
    =& C \cdot \min \left(\exp [-\beta \Phi(\mathbf{w}')], \exp [-\beta \Phi(\mathbf{w})] \right) \\
    =& \frac{\exp [-\beta \Phi(\mathbf{w}')]}{\sum_{\bar{\mathbf{w}} \in \mathcal{W}} \exp [-\beta \Phi(\bar{\mathbf{w}})]} \cdot \frac{1}{N|\mathcal{W}_i|} \cdot \min\left(\frac{\exp(-\beta \Phi(\mathbf{w}))}{\exp(-\beta \Phi(\mathbf{w}'))}, 1\right) \\
    =& \pi^*_{\mathbf{w}'} \cdot q_{\mathbf{w}', \mathbf{w}} \cdot a_{\mathbf{w}', \mathbf{w}},
\end{split}
\end{equation}
where $C \triangleq ({{N|\mathcal{W}_i| \cdot \sum_{\bar{\mathbf{w}} \in \mathcal{W}} \exp [-\beta \Phi(\bar{\mathbf{w}})]}})^{-1}$ is a constant. 

Similar to the design of MA-SCCA, we can reformulate the acceptance probability so that it uses only local information. Formally, by applying (\ref{eq: potential_cost}), we have
\begin{equation}\label{eq: accept_prob2}
    a_{(\mathbf{w}_i, \mathbf{w}_{-i}), (\mathbf{w}'_i, \mathbf{w}_{-i})} = \min\left(\frac{\exp(-2\beta \lambda_i \gamma_{\text{(VM)}}^{-F_i} c_i(\mathbf{w}'_i, \mathbf{w}_{-i}))}{\exp(-2\beta \lambda_i \gamma_{\text{(VM)}}^{-F_i} c_i(\mathbf{w}_i, \mathbf{w}_{-i}))}, 1\right).
\end{equation}
    
\textbf{Remarks:} The pseudocode of MH-SCCA differs with MA-SCCA only in line $11$ of {Algorithm \ref{alg}}. Instead of taking a strategy according to the probability distribution (\ref{eq: trans_prob_alg2}), under MH-SCCA, player $i$ randomly uniformly picks a strategy and accepts it with probability (\ref{eq: accept_prob2}). 
Such a minor change significantly reduces the overall computational complexity. The reason is that, in each iteration, MA-SCCA requires the costs of all states that the chosen player $i$ can reach by changing its own strategy, while MH-SCCA requires only the cost of the current state and the proposal state. As a result, such improvements would lead to a faster convergence rate in practice, which is verified by simulations in Section \ref{sec: numerical results}. 

\subsection{Performance Analysis}
We denote $\pi^t$ as the probability distribution over all states in $\mathcal{W}$ at the $t$-th iteration. 
We adopt the \textit{total variation distance}  \cite{diaconis1991geometric} as the metric to characterize the difference between $\pi^t$ and the desired distribution $\pi^*$. Formally, the total variance distance between $\pi^t$ and $\pi^*$ is given by
\begin{equation}\label{def: tv}
    \Vert \pi^t - \pi^* \Vert_{TV} \triangleq \frac{1}{2} \sum_{\mathbf{w} \in \mathcal{W}} \vert \pi^t_{\mathbf{w}} - \pi^*_{\mathbf{w}} \vert. 
\end{equation}
Then we define the mixing time (convergence time) of the Markov chain as follows 
\begin{equation}\label{eq: def_tau}
    t_{\text{mix}}(\varepsilon) \triangleq \min \{t \geq 0:  \Vert \pi^t - \pi^* \Vert_{TV} \leq \varepsilon\},
\end{equation}
which measures the minimum number of iterations required to bound the total variation distance within a small value $\varepsilon$. 
Regarding MA-SCCA and MH-SCCA, we have the following theorem on their mixing times.
\begin{theorem}\label{theorem: performance}
	\textit{
	The mixing times of MA-SCCA and MH-SCCA are upper-bounded as follows  
	\begin{equation}\label{eq: perf}
		t_{\text{mix}}(\varepsilon) \leq \left \lceil \frac{\log{\frac{1}{2\varepsilon}}+\frac{1}{2} FN \log{M}+\frac{1}{2} \beta D}{-\log{(1-\frac{1}{2 M^{2F(N+1)}}\exp{[-4\beta D]})}} \right \rceil,
	\end{equation}
    where $D$ is the difference between the maximum and minimum potential values and defined as
    \begin{equation}\label{def: D}
        D \triangleq \max_{\mathbf{w} \in \mathcal{W}} \Phi(\mathbf{w}) - \min_{\mathbf{w} \in \mathcal{W}} \Phi(\mathbf{w}).
    \end{equation}
	}
\end{theorem}
The proof is relegated to Appendix-D.

Theorem \ref{theorem: performance} shows that the upper bound of the mixing time increases proportionally to the number of players $N$, the number of servers $M$, and the number of VNF types $F$. 
Moreover, we also find that increasing the value of $\beta$ will not only reduce the approximation gap (see Theorem \ref{theorem: gap}), but also prolong the mixing time. 
In practice, it is important to make a trade-off between optimality and convergence based on the system design objective.
Note that, although MH-SCCA and MA-SCCA are shown to have the same upper bound of mixing times, MH-SCCA often converges faster than MA-SCCA in practice. We will further verify this in Section \ref{sec: numerical results}.

\section{Learning-aided Scheme Design} \label{sec: learning-aided algo design}
As comparisons to MA-SCCA and MH-SCCA, in this section, we propose two novel learning-aided service chain composition schemes. The two scheme designs are based on deep reinforcement learning (DRL) methods and Monte Carlo tree search (MCTS) methods, respectively.

\subsection{DRL-based Service Chain Composition}  \label{subsection: drl}
\begin{figure}[!t]
    \centering
    \includegraphics [scale=0.35] {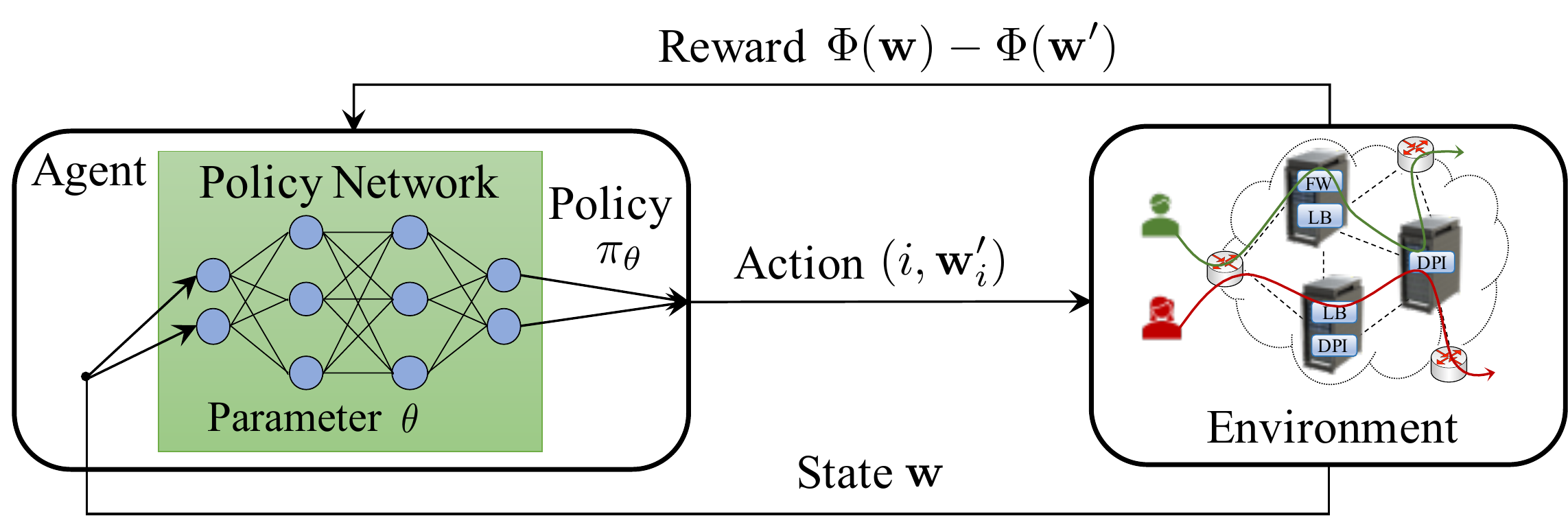}
    \centering 
    \caption{Overview of DRL-SCCA design.}
    \label{drl}
\end{figure}

In this subsection, we first review some basics of DRL, then show the design of our DRL-based Service Chain Composition Algorithm (DRL-SCCA), as visualized in Figure \ref{drl}.

Over the past few years, deep reinforcement learning (DRL) techniques have been widely adopted in various fields to solve complicated decision making problems and achieved encouraging results \cite{mnih2015human} \cite{nair2018visual}.
Under a typical DRL setting, there is an agent continuously interacting with its environment over iterations. 
At the beginning of each iteration $t$, the agent observes the environment state $s_t$. Meanwhile, the agent maintains a policy $\boldsymbol{\pi}$ that decides the action $a_t$ based on the observed state $s_t$.
Depending on the agent's action, the environment evolves to the next state $s_{t+1}$ and reveals a reward signal $r_t$ to the agent. The agent then exploits such information to improve its policy.
In DRL, the policy is usually implemented as a deep neural network (DNN) with parameters $\theta$. 
The long-term goal of the agent is to maximize the expectation of its cumulative rewards, denoted by $\mathbb{E}[\sum_{t=0}^{\infty} \gamma^{t} r_t]$, where the constant $\gamma \in [0, 1]$ is a discounting factor that weighs the impact of future rewards.  

To apply DRL to solve service chain composition problem, the key lies in how we cast service chain composition into a decision making process.
Following the similar idea in MA-SCCA and MH-SCCA, we consider the service chain composition as an iterative process. Given a randomly initialized state $\mathbf{w}$, the agent conducts a series of iterations to update strategies for players. In each iteration, the agent only picks one player to update its strategy. Accordingly, the agent will receive a reward that indicates the benefit resulting from such update. Through such a series of updates, the agent aims to achieve the state with the minimum potential value.
In the following, we illustrate the detailed design of the state representation, agent action, reward design, and the policy network of the agent, respectively. The pseudocode of DRL-SCCA is given in Algorithm \ref{algo: DRL-SCCA}.

\begin{algorithm}[!b]
    \caption{DRL-based Service Chain Composition Algorithm (DRL-SCCA)}
    \begin{algorithmic}[1] \label{algo: DRL-SCCA}
        \renewcommand{\algorithmicrequire}{\textbf{Input:}}
        \renewcommand{\algorithmicensure}{\textbf{Output:}}
        \REQUIRE $N$ players, each player $i$ with a strategy set $\mathcal{W}_{i}$.
        \ENSURE Service chain composition decision $\textbf{w}$ 
        \STATE \textbf{for} each player $i = 1, 2, \dots, N,$ \textbf{do}
        \STATE $~~$ Select a strategy $\mathbf{w}_i$ uniformly randomly from $\mathcal{W}_{i}$.
        \STATE Initialize observed state $\mathbf{w} \leftarrow (\mathbf{w}_1, \dots, \mathbf{w}_i, \dots, \mathbf{w}_N)$.
        \STATE \textbf{for} each iteration \textbf{do}
        \STATE $~~$ 
        Acquire action $(i, \mathbf{w}_i') \leftarrow \boldsymbol{\pi}_{\theta}(\mathbf{w})$.
        \STATE $~~$ Update $\mathbf{w} \leftarrow (\mathbf{w}_1, \dots, \mathbf{w}'_i, \dots, \mathbf{w}_N)$.
        \STATE \textbf{return} $\mathbf{w}$ 
    \end{algorithmic}
\end{algorithm}
    
\textbf{State Representation:} 
The state observed by the agent contains all players' service chain composition strategies, \textit{i.e.}, $\mathbf{w} = (\mathbf{w}_1, \cdots, \mathbf{w}_N)$.

\textbf{Agent Action:}
We define each action taken by the agent as a tuple. For example, the action $(i, \mathbf{w}'_i)$ implies that the agent decides to update the strategy of player $i$ as $\mathbf{w}'_i$. Consequently, the state evolves to $\mathbf{w}' = (\mathbf{w}_1, \cdots, \mathbf{w}'_i, \cdots, \mathbf{w}_N)$.

\textbf{Reward Design:}
We define the reward as 
$\Phi(\mathbf{w}) - \Phi(\mathbf{w}')$, \textit{i.e.},
the difference between the potential value before and after the agent takes the action. 
Such a reward setting unifies the objective of maximizing cumulative rewards and minimizing the potential value.
Following (\ref{eq: potential_cost}), the reward can be reformulated as $2\lambda_i \gamma_{\text{(VM)}}^{-F_i} \left( c_i(\mathbf{w}) - c_i(\mathbf{w}') \right)$.

\begin{figure}[!t]
    \centering
    \includegraphics [scale=0.35] {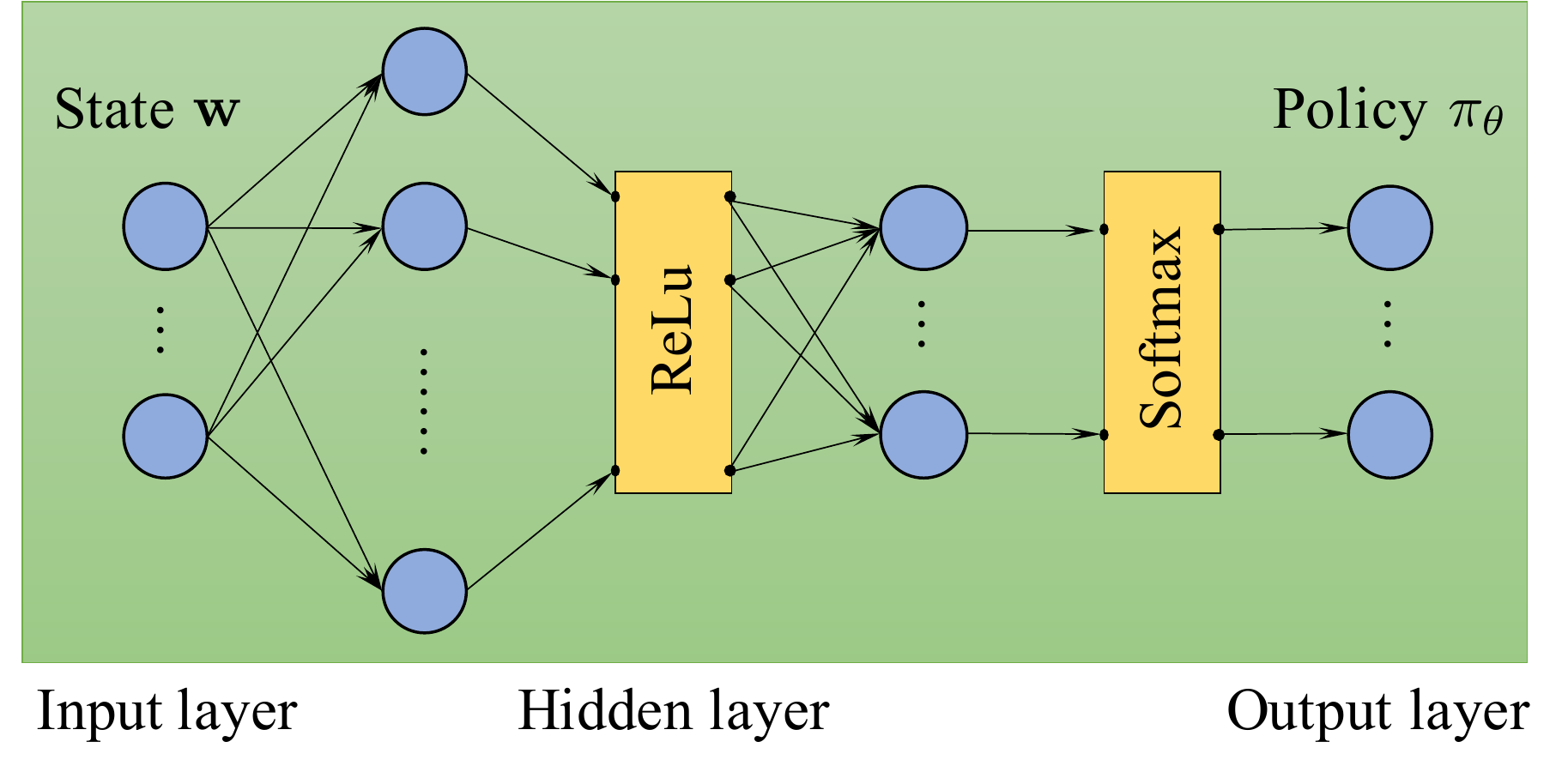}
    \centering 
    \caption{Design of the policy network.}
    \label{dnn}
\end{figure}

\textbf{Policy Network:} 
We design the agent's policy network as a forward neural network (FNN) with one hidden layer of dimension $2048$, followed by a ReLu activation function, as visualized in Figure \ref{dnn}.
The policy network takes the observed state $\mathbf{w}$ as input and the probability distribution $\boldsymbol{\pi}_{\theta}(\cdot \vert \mathbf{w})$ as output.
We note that such a policy network design requires the number of players $N$, the number of servers $M$, and the number of VNF types $F$ to be fixed. 
The change in any of such numbers would require a reconstruction and re-training of the policy network. 
Nevertheless, our design can serve as a basis for more ingenious policy network design in the future.

We adopt the policy gradient (PG) method \cite{sutton2000policy} to train the policy network, aiming to maximize the cumulative rewards. 
The key idea of PG is that if an action results in more rewards, the agent will adjust its policy network so that next time when observing the same state, it will be more likely to adopt such an action. 
Likewise, the agent will reduce its willingness to take the same action next time when observing the same state if the action leads to few rewards. 

\subsection{MCTS-based Service Chain Composition} \label{subsection: mcts}
In fact, the search for the optimal solution of problem (\ref{eq: problem_initial}) can also be viewed as a \textit{sequential decision making} process. 
Particularly, in such a process, a solution can be attained by determining the strategy for each player successively. 
Therefore, problem (\ref{eq: problem_initial}) is equivalent to a \textit{tree-search} problem with respect to a decision tree. Each \textit{node} in the tree corresponds to an \textit{action selection state} and each directed edge from the node in the $l$-th layer to the child node in the $(l+1)$-th layer represents a particular strategy selection for the $l$-th player.
For example, the root node in the tree represents an \textit{empty state}, in which no player chooses any strategy. 
Each edge from the root node represents a possible strategy selection for the first player. Accordingly, the resulting child node corresponds to the state after the selection. 
Consequently, each leaf node in the tree corresponds to a feasible service chain composition decision.
In this way, solving problem (\ref{eq: problem_initial}) is equivalent to searching for a path in the tree that leads to the leaf node with the minimum potential value. 
Recall that problem (\ref{eq: problem_initial}) is in general $\mathcal{NP}$-hard with an enormous search space size. Consequently, this implies a huge number of leaf nodes in the decision tree, which makes a thorough search over all possible paths impractical.

\begin{algorithm}[!b]
     \caption{MCTS-based Service Chain Composition Algorithm (MCTS-SCCA)}
     \begin{algorithmic}[1]
     \label{algo: MCTS-SCCA}
     \renewcommand{\algorithmicrequire}{\textbf{Input:}}
     \renewcommand{\algorithmicensure}{\textbf{Output:}}
     \REQUIRE ${N}$ players, each player $i$ with a strategy set $\mathcal{W}_{i}$.
     \ENSURE Service chain composition decision  $\textbf{w}$ 
     \STATE Initialize set $\mathcal{Z} \leftarrow \emptyset$.
     \STATE \textbf{for} each player $i = 1, 2, \dots, N,$ \textbf{do}
     \STATE $~~$ 
         $\mathbf{w}_i \leftarrow \text{SELECT-STRATEGY}( \mathcal{Z}, i)$
     \STATE $~~$ Update $\mathcal{Z} \leftarrow \mathcal{Z} \bigcup \{ (i, \mathbf{w}_i) \}$.
    \STATE \textbf{return} $\mathbf{w}$
    \end{algorithmic}
\end{algorithm}

In recent years, Monte Carlo tree search (MCTS) methods have been proven effective to solve large-scale complicated tree-search problems \cite{browne2012survey}. 
The key idea of MCTS is that, instead of building the whole tree in one shot, the search should be carried out by constructing only part of the tree incrementally. Specifically, MCTS maintains only those branches that are the most promising to lead to the best leaf nodes based on the estimation from a number of \textit{random sampling}. Such a randomized scheme, if properly applied, can achieve a decent balance between tractability and optimality. 
Inspired by such an idea, we propose an MCTS-based Service Chain Composition Algorithm (MCTS-SCCA), that aims to search the leaf node in the decision tree with the optimal solution of the original problem (\ref{eq: problem_initial}). 
We show the pseudocode of MCTS-SCCA in Algorithm \ref{algo: MCTS-SCCA}.

\begin{figure}[!t]
    \centering
    \includegraphics [scale=0.32] {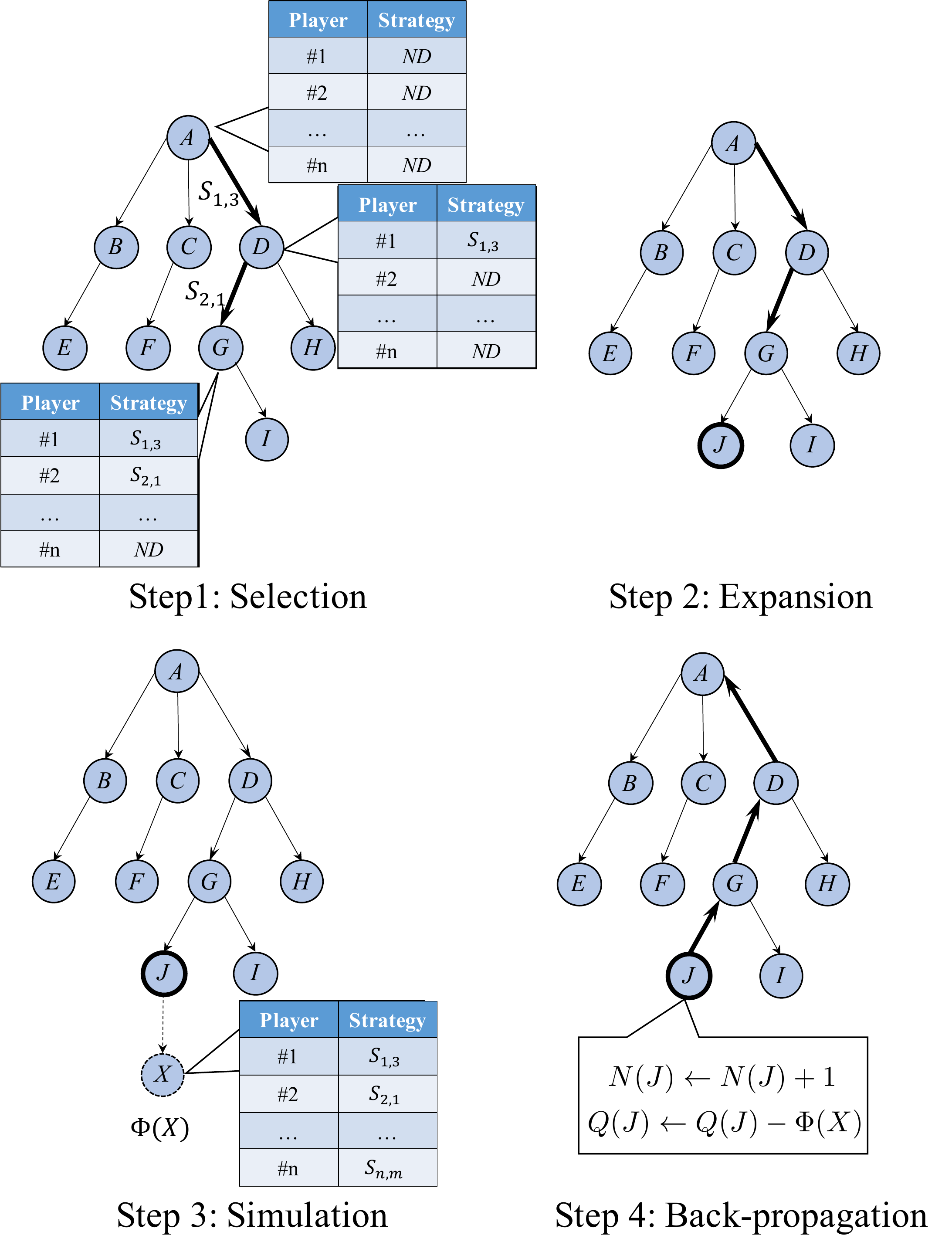}
    \caption{One-round sampling in MCTS-SCCA, where \textit{ND} stands for ``not determined". $S_{i, k}$ is the $k$-th strategy in the strategy set of player $i$. The leaf node $X$ corresponds to one feasible service chain composition decision $\mathbf{w}$.}
    \label{MCTS-SCCA}
\end{figure}

\subsection{Design of MCTS-SCCA}
To apply MCTS to problem (\ref{eq: problem_initial}), we associate each node $\eta$ in the decision tree with four variables: the number of times node $\eta$ has been visited $N(\eta)$, its received cumulative reward $Q(\eta)$, the set of all child nodes $\mathcal{C}(\eta)$, and the set of unvisited child nodes $\mathcal{C}_{\text{untried}}(\eta)$. 
Basically, MCTS-SCCA successively selects the most promising strategy for each player with rounds of random sampling. 
The assignment state, denoted by $\mathcal{Z}$, includes the pairs of players and their strategies.
Upon the strategy selection for player $i$, $\mathcal{Z}$ contains the strategies of players $1, \dots, i-1$. We define $n(\mathcal{Z})$ as the current node. For example, in Figure \ref{MCTS-SCCA} step 1, we have $\mathcal{Z} = \{(1, S_{1, 3}), (2, S_{2, 1})\}$, and the current node $n(\mathcal{Z})$ is $G$. MCTS-SCCA proceeds as shown in Algorithm \ref{algo: mcs-subfunc}. 

Starting from node $n(\mathcal{Z})$ for the assignment state $\mathcal{Z}$, MCTS-SCCA will take a series of sampling rounds (lines $3$-$10$ in Algorithm \ref{algo: mcs-subfunc}) to estimate the ``advantage" of choosing each strategy $\mathbf{w}_i$ for player $i$. Figure \ref{MCTS-SCCA} shows how MCTS-SCCA proceeds within one round of sampling.

\begin{algorithm}[!t]
	\caption{Sub-functions for MCTS-SCCA}
	\begin{algorithmic}[1]
	\label{algo: mcs-subfunc}
	\STATE \textbf{function} SELECT-STRATEGY($\mathcal{Z}$, $i$) \\
	\STATE $~~~~$ Initialize $k \leftarrow 0$.
	\STATE $~~~~$ \textbf{while} $k <$ MAX\_SAMPLE\_NUM \textbf{do}
	\STATE $~~~~~~~~$ $(n(\mathcal{Z}'), i') \leftarrow \text{TRAVERSE}(\mathcal{Z}, i)$
	\STATE $~~~~~~~~$ \textbf{If} $n(\mathcal{Z}')$ is not a leaf:
	\STATE $~~~~~~~~~~~~$ $\Phi \leftarrow \text{SIMULATE}(\mathcal{Z}', i')$
	\STATE $~~~~~~~~$ \textbf{Else}:
	\STATE $~~~~~~~~~~~~$ $\Phi \leftarrow \Phi(\mathcal{Z}')$
	\STATE $~~~~~~~~$ $\text{BACK\_PROP}(\mathcal{Z}', \Phi)$ 
	\STATE $~~~~~~~~$ $k \leftarrow k + 1$
	\STATE $~~~~$ $\eta_{\text{best}} \leftarrow \text{BEST\_CHILD}(n(\mathcal{Z}), 0)$
 	\STATE $~~~~$ \textbf{return} strategy $\mathbf{w}_i$ that leads from $n(\mathcal{Z})$ to $\eta_{\text{best}}$
 	\STATE
 	\STATE \textbf{function} TRAVERSE($\mathcal{Z}$, $i$)
 	\STATE $~~~~$ Initialize $\mathcal{Z'} \leftarrow \mathcal{Z}$, $\eta \leftarrow n(\mathcal{Z}')$ and $i' \leftarrow i$.
 	\STATE $~~~~$ \textbf{while} $\eta$ is not a leaf \textbf{do}
 	\STATE $~~~~~~~~$ \textbf{if} $\mathcal{C}_{\text{untried}}(\eta) \neq \emptyset$ \textbf{then}
 	\STATE $~~~~~~~~~~~~$ Pick $\eta' \in \mathcal{C}_{\text{untried}}(\eta)$ uniformly at random.
 	\STATE $~~~~~~~~~~~~$ Set $\mathcal{Z}' \leftarrow \mathcal{Z}' \cup \{ \eta' \}$.
 	\STATE $~~~~~~~~~~~~$ \textbf{return} $(n(\mathcal{Z}'), i')$
 	\STATE $~~~~~~~~$ \textbf{else}
 	\STATE $~~~~~~~~~~~~$ Set $\eta \leftarrow \text{BEST\_CHILD}(\eta, \sqrt{2})$.
 	\STATE $~~~~~~~~~~~~$ Set $i' \leftarrow i' + 1$.
 	\STATE $~~~~$ \textbf{return} $(\eta, i')$
 	\STATE
 	\STATE \textbf{function} BEST\_CHILD($\eta$, $\omega$)
 	\STATE $~~~~$ \textbf{return} $\underset{\eta' \in \mathcal{C}(\eta)}{\arg\max}\,\, 
 	\frac{Q(\eta')}{N(\eta')} + \omega \sqrt{\frac{
 		2 \ln N(\eta)}{N(\eta')}}$ 
 	\STATE
 	\STATE \textbf{function} SIMULATE($\mathcal{Z}$, $i$)
	 \STATE $~~~~$ Initialize $\mathcal{Z}' \leftarrow \mathcal{Z}$, $i' \leftarrow i$.
 	\STATE $~~~~$ \textbf{while} $i' < n$ \textbf{do}
 	\STATE $~~~~~~~~$ Pick a strategy $\mathbf{w} \in \mathcal{W}_{i'+1}$ uniformly at random.
 	\STATE $~~~~~~~~$ Set $i' \leftarrow i' + 1$.
 	\STATE $~~~~~~~~$ Set $\mathcal{Z}' \leftarrow \mathcal{Z}' \cup \{ (i', \mathbf{w}) \}$.
 	\STATE $~~~~$ \textbf{return} $\Phi(\mathcal{Z}')$
 	\STATE 
 	\STATE \textbf{function} BACK\_PROP($\eta$, $\Phi$)
	 \STATE $~~~~$ \textbf{while} $\eta$ is not the root node \textbf{do}
 	\STATE $~~~~~~~~$ Set $N(\eta) \leftarrow N(\eta) + 1$ and $Q(\eta) \leftarrow Q(\eta) - \Phi$.
 	\STATE $~~~~~~~~$ Set $\eta \leftarrow$ parent($\eta$).
\end{algorithmic}
\end{algorithm}

At the beginning of each round of sampling, MCTS-SCCA maintains a partial tree which contains nodes that have been marked as visited in previous rounds. Then MCTS-SCCA takes the following four steps to carry out the sampling. 
\subsubsection{Selection}
MCTS-SCCA first takes a path from the root node over the partially constructed tree. In such a process, the selection of child nodes must address the \textit{exploitation-exploration} dilemma: should MCTS-SCCA choose the child node with the highest empirical mean, or should it explore those rarely visited child nodes that may possibly result in even better performance? 
	To this end, MCTS-SCCA maintains an upper-confidence-bound (UCB) value for each node (line $27$ in Algorithm \ref{algo: mcs-subfunc}), which equals the weighted sum of an exploration term (the average reward through that node) and an exploitation term. Note that, in MCTS-SCCA, we set the weight $\omega$ for the exploration term as $\sqrt{2}$ which is a widely adopted empirical choice  \cite{browne2012survey}. Intuitively, the second term will be large if a node has been visited rarely given that its parent node has been visited for a large number of times. The traversal stops by an anchor node, which is either a leaf node or some node with any unvisited child node. 
	If stopped by a leaf node, then directly jumps to step \textit{3)} . 

\subsubsection{Expansion} 
If stopped by a non-leaf node with unvisited child nodes, MCTS-SCCA picks one of the unvisited child nodes of the anchor node randomly, then marks the selected child node as visited, and adds it to the partial tree.   

\subsubsection{Simulation} 
Given the selected child node with the corresponding assignment state, MCTS-SCCA will take a random path leading to some leaf node. In other words, MCTS-SCCA will uniformly randomly pick a strategy for each unassigned player and yield a feasible decision $\mathbf{w}$. 

\subsubsection{Back-propagation} 
MCTS-SCCA will use the potential value $\Phi(\mathbf{w})$ of the eventually reached leaf node to update the values of $N(\eta)$ and $Q(\eta)$ for each node $\eta$ along the path. 

With the updated estimated rewards based on rounds of sampling, MCTS-SCCA chooses the current root node's best child node with the maximum empirical mean reward (line $11$ in Algorithm \ref{algo: mcs-subfunc} with parameter $\omega = 0$). Then MCTS-SCCA sets the chosen child node as the new root and conducts a new series of random sampling over the new root's tree. Such a process will be repeated until the chosen child node is a leaf node. MCTS-SCCA will output the corresponding state as the final service chain composition decision.

\subsection{Discussion}
Compared to centralized schemes DRL-SCCA and MCTS-SCCA, the distributed implementations of MA-SCCA and MH-SCCA conduce to better adaptability. 
Though limited to centralized implementations without provable performance guarantee, DRL-SCCA and MCTS-SCCA take other advantages. 
For DRL-SCCA, it solves problem (\ref{eq: problem_initial}) in a data-driven and \textit{model-free} fashion. 
In other words, the design of DRL-SCCA does not require specific domain knowledge about the problem itself. The crux lies in the design of reward signals, the representations of states and agent actions, respectively. However, DRL-SCCA requires a large amount of time for pre-training before applied to the practical system. 
For MCTS-SCCA, it enjoys a low computational complexity that is proportional to the number of players and the number of samples (MAX\_SAMPLE\_NUM). However, insufficient sampling may result in solutions that are far from optimum.

\section{Numerical Results}  \label{sec: numerical results}
In this section, we evaluate the performance of MA-SCCA, MH-SCCA, DRL-SCCA, and MCTS-SCCA under various scenarios.
In the following, we first illustrate the basic settings and the baseline scheme adopted in our simulation, then we show the simulation results and the corresponding analysis. 

\subsection{Simulation Settings}
We adopt the following settings in our simulation by default unless otherwise specified. All simulation results are obtained by averaging the results from $100$ runs. 
    
\textbf{Basic Settings:}
We consider an NFV system equipped with five servers. Each server holds three instances, corresponding to three commonly adopted VNFs, \textit{viz.} Firewall (FW), Load Balancer (LB), and Intrusion Detection System (IDS). 
We set the latency between any gateway router $r\in \mathcal{R}$ and any VM $v \in \mathcal{V}$, and the latency between any two VMs within the same server uniformly as one unit of time. The latency between any other pair of VMs is uniformly sampled from the interval $[2, 6]$.   
In the meantime, we consider ten players, each with a traffic rate of $5$Mbps and a failure cost of $\Omega=1000$ units. 
All players are assumed to subscribe the same service chain, \textit{i.e.}, FW $\rightarrow$ LB $\rightarrow$ IDS.    
Besides, the VM survival probability and user survival probability are both set to $0.9$. 
By viewing latency reduction and congestion mitigation as equally important, we set $\alpha=1$. 
For MA-SCCA and MH-SCCA, we set $\beta=0.1$.
Under such settings, there are multiple Nash equilibria with sub-optimal potential values, and the optimal potential value should be about $1575$ units. 
    
\textbf{Baseline:}
 We adopt the Unilateral Service Chain Selection (USCS) scheme  \cite{d2017exploiting} as a baseline. The implementation of USCS is as follows. Initially, each player randomly picks one strategy from its strategy space. Then players take turns to make strategy changes. In each iteration, the chosen player will change its strategy to the one that minimizes its own cost. Such a procedure ends if no players are willing to change its strategy.\footnote{Note that MA-SCCA will degrade to USCS by setting  $\beta \rightarrow \infty$.}
Similar to MH-SCCA, USCS is also a distributed scheme because each player can act individually. 
However, the complexity within each iteration of USCS is very high because it requires a thorough search over the strategy space to find the strategy with the minimum cost. 
In fact, USCS is some kind of ``best response" scheme. Such schemes usually run  fast, but may fall into some local optima, inducing inferior system performances.
 
\textbf{Scheme Settings:}
We assume the NFV system proceeds on a time-slot basis. 
For MA-SCCA, MH-SCCA, DRL-SCCA, and USCS, they run one iteration within each time slot.
Considering the difference among their computational complexities, we set the time slot length as one second for USCS, MA-SCCA and DRL-SCCA, and ten milliseconds for MH-SCCA. 
Particularly, the mean of the Poisson clock is set as $100$ and $1$ millisecond for MA-SCCA and MH-SCCA, respectively.
For DRL-SCCA, we train its policy network by setting the learning rate as $0.0001$.
For MCTS-SCCA, we do not consider any time slot settings since the service chain composition is completed in one shot and remains unchanged all the time. 

\subsection{Performance Comparison}
To evaluate the performance of the proposed schemes, we consider two different metrics: the \textit{potential value} in (\ref{eq: potential}) and the \textit{weighted average cost} which is defined as
\begin{equation}\label{def_C_hat}
    \bar{C} \triangleq \frac{1}{N} \sum_{i\in \mathcal{N}} \lambda_i c_i(\mathbf{w}_i, \mathbf{w}_{-i}).
\end{equation}
Under the default simulation settings, the minimum weighted average cost should be about $1826$ units.

\begin{figure}[!t]
    \centering
    \subfloat[Potential value.]{
    \label{subfig: comparison_potential}
    \includegraphics[scale=.15]{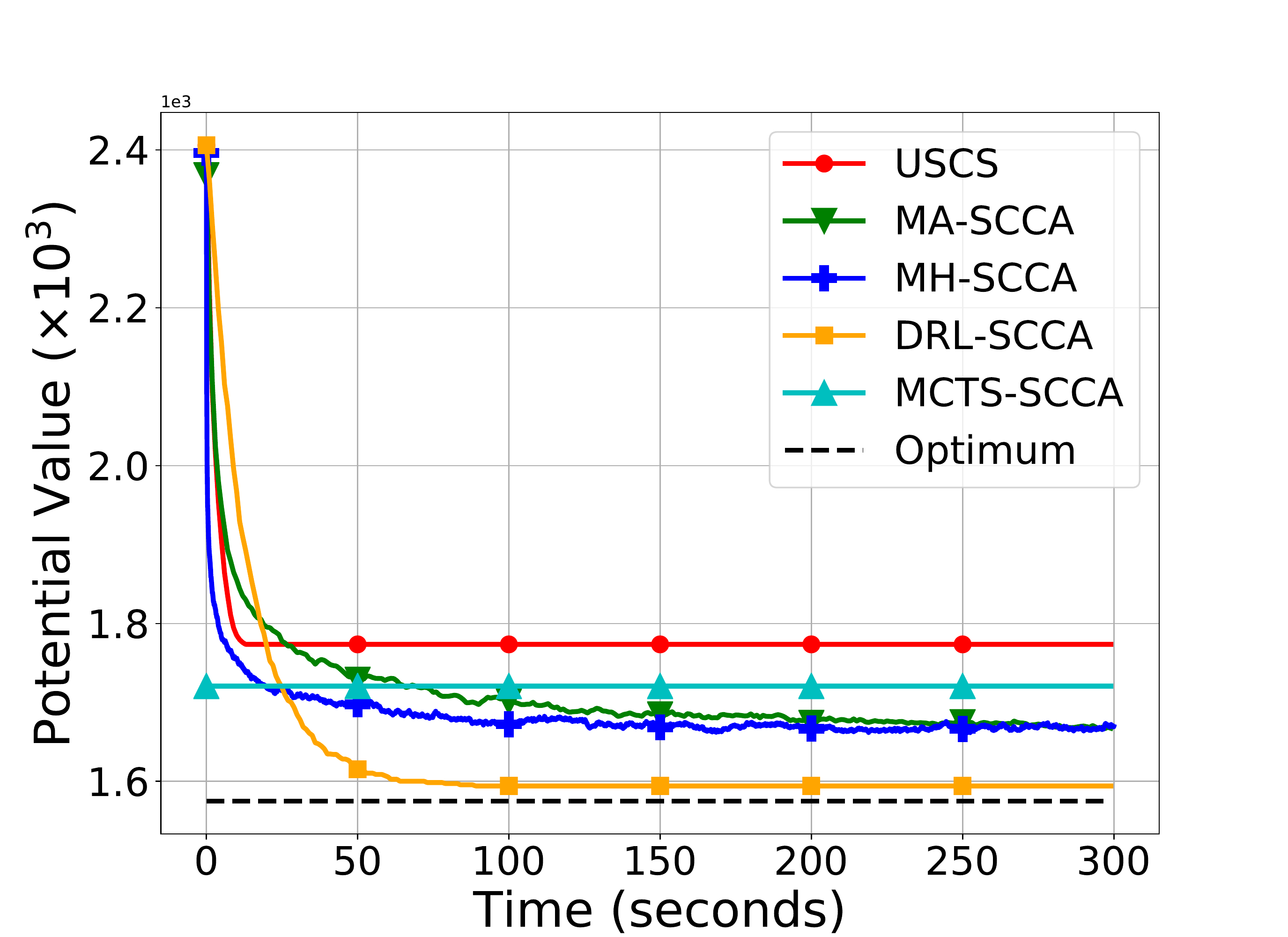}}
    \hfil
    \subfloat[Weighted average cost.]{
    \label{subfig: comparison_cost}
    \includegraphics[scale=.15]{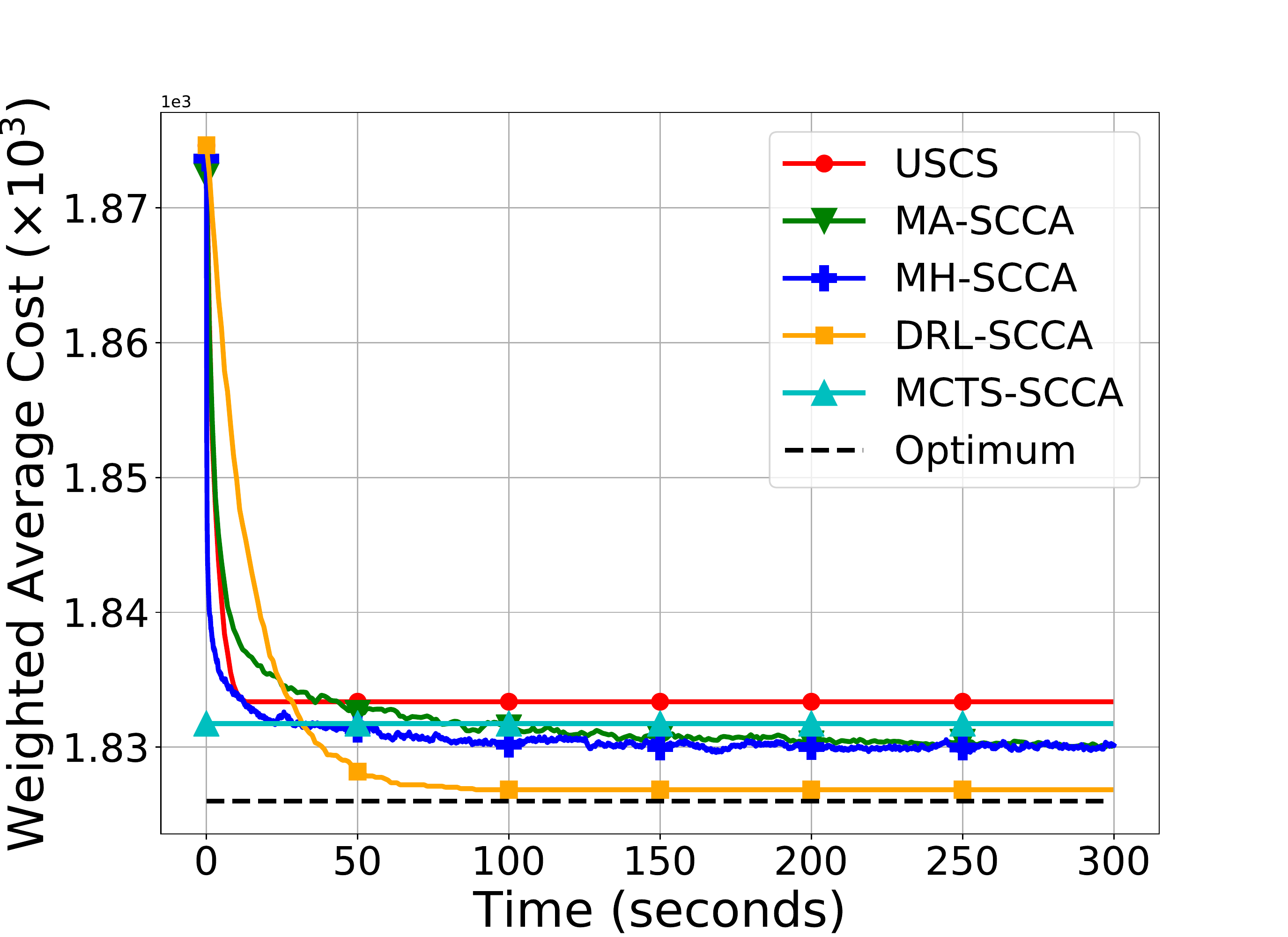}}
    \caption{Comparison of USCS, MA-SCCA, MH-SCCA, DRL-SCCA, and MCTS-SCCA in terms of (a) potential value and (b) weighted average cost.}
    \label{fig: comparison}
\end{figure}

Figure \ref{fig: comparison} shows the comparison results of USCS, MA-SCCA, MH-SCCA, DRL-SCCA, and MCTS-SCCA in terms of (a) potential value and (b) weighted average cost. The pre-training times for DRL-SCCA and MCTS-SCCA are half an hour and three minutes, respectively. 
We have the following observations:  
1) The curves in Figures \ref{fig: comparison}(a) and \ref{fig: comparison}(b) are very similar, which verifies the proportional relationship between the local change of cost and the global change of potential value as shown in (\ref{eq: potential_cost}). 
2) The curves of all schemes except MCTS-SCCA start from the same point because all these schemes begin with a random initialization of all players' strategies;
3) USCS converges very fast, but it only reaches some local optimum, and stays unchanged. 
4) Compared with USCS, MA-SCCA achieves both a smaller potential value and a lower weighted average cost. 
5) Compared with MA-SCCA, the curves of MH-SCCA converges faster, because MH-SCCA has a lower computational complexity. Nonetheless, after a period of time, the two curves coincide eventually. The reason is that, with the same value of $\beta = 0.1$, MA-SCCA and MH-SCCA result in the same stationary distribution (\ref{eq: opt_prob}), resulting in the same expected potential value and expected weighted average cost.  
6) DRL-SCCA outperforms all other schemes in terms of both potential value and weighted average cost. The reason is that it has been trained in advance for about half an hour, and can guide the players towards the optimal system performance at runtime. 
7) Using only three minutes to make decisions in advance, MCTS-SCCA reaches good performance which is in between USCS and MH-SCCA. 

Both DRL-SCCA and MCTS-SCCA are static schemes, which are not suitable for NFV systems with frequent dynamics.
In addition, MA-SCCA and MH-SCCA are very similar, but the complexity of MA-SCCA is much more higher. 
Based on the above considerations, in the following simulations, we concentrate on only two schemes: USCS and MH-SCCA.
    
\begin{figure}[!t]
    \centering
    \subfloat[Potential value.]{
    \label{subfig: beta_potential}
    \includegraphics[scale=.15]{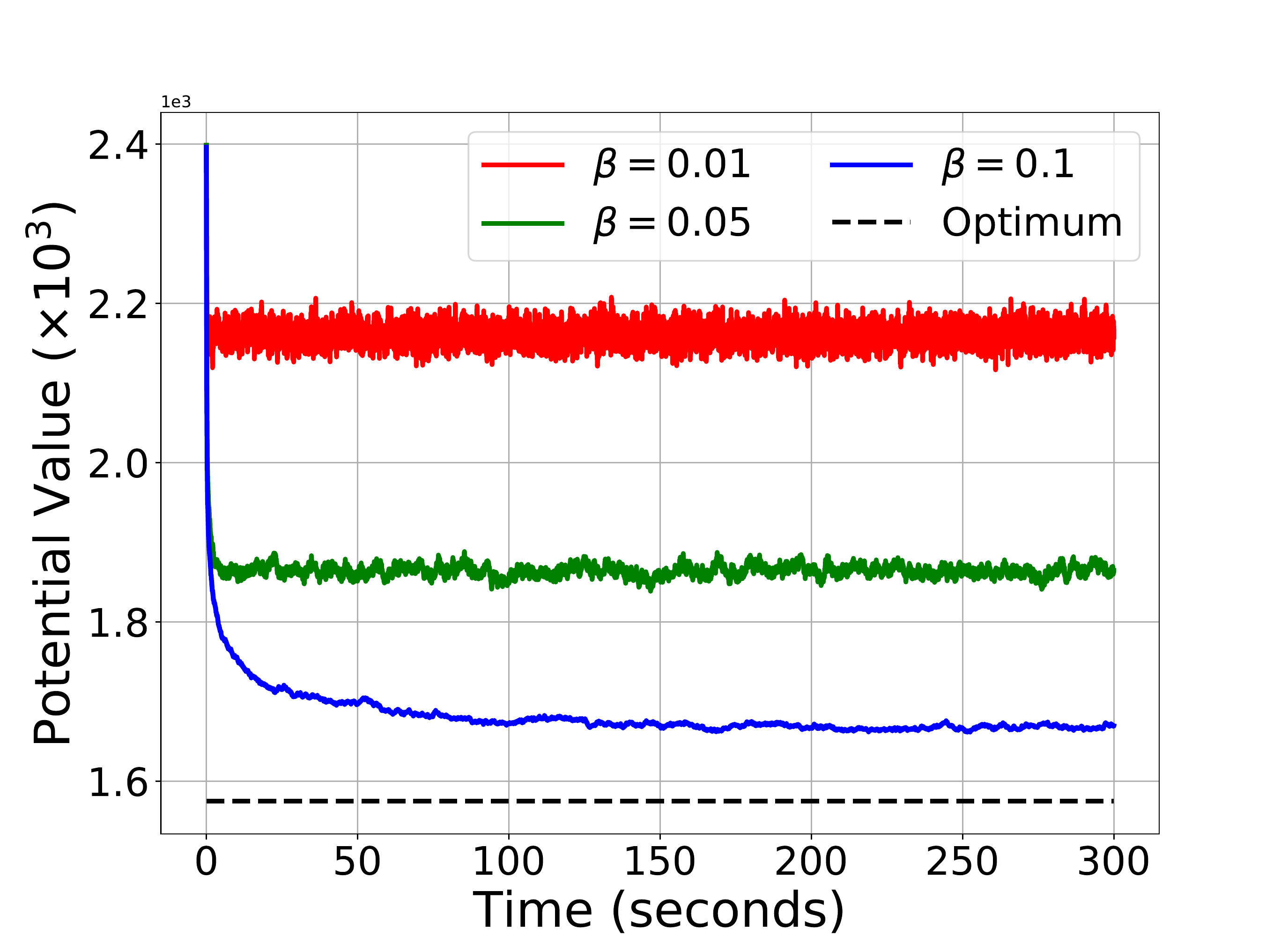}}
    \hfil
    \subfloat[Weighted average cost.]{
    \label{subfig: beta_cost}
    \includegraphics[scale=.15]{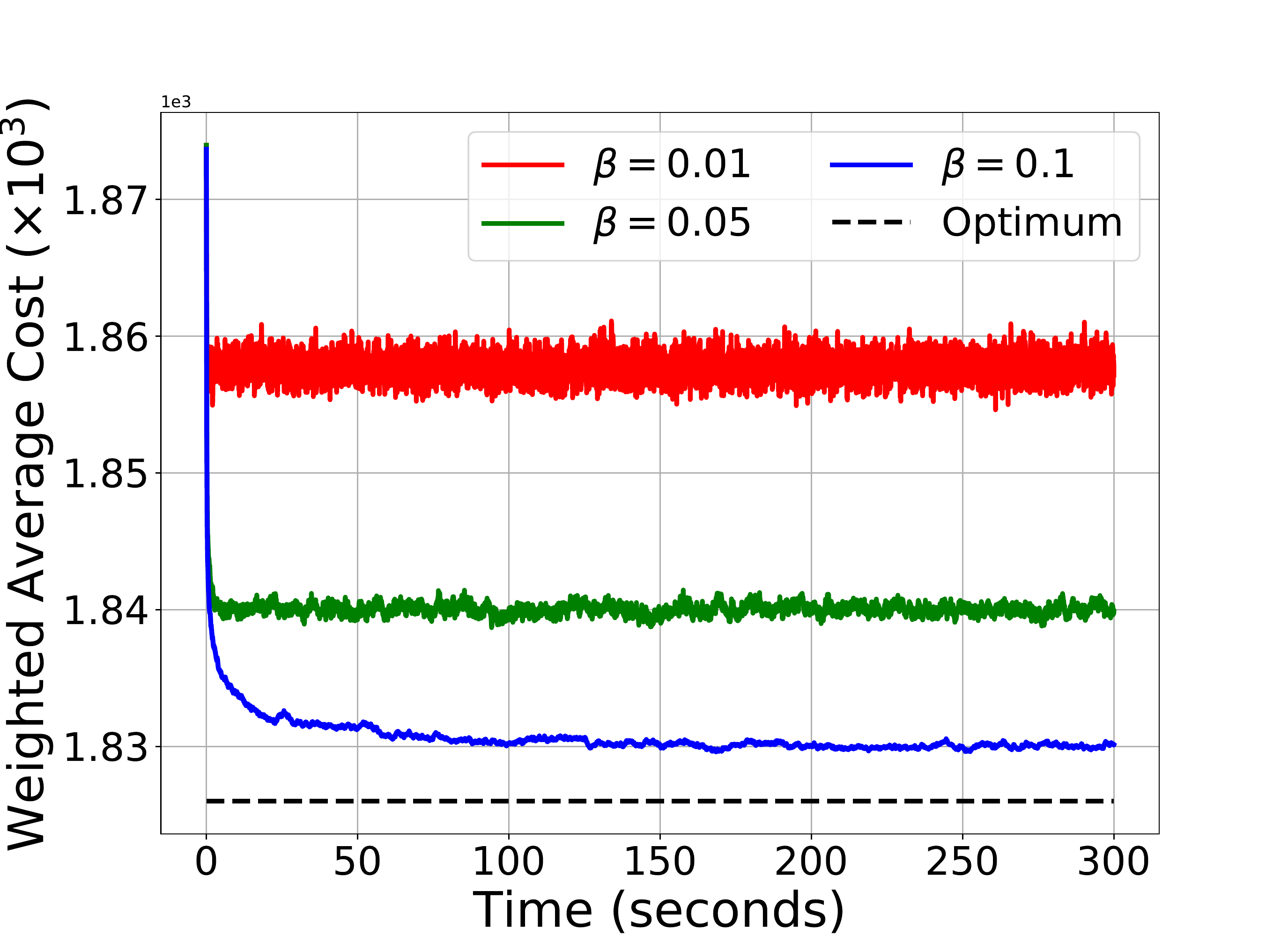}}
    \caption{Performance of MH-SCCA under various values of $\beta$ in terms of (a) potential value and (b) weighted average cost.}
    \label{fig: beta}
\end{figure}

\subsection{Evaluation of USCS and MH-SCCA}
Figure \ref{fig: beta} presents the results of MH-SCCA under different values of $\beta$. We find that both the potential value and the weighted average cost \textit{after convergence} decrease as the value of $\beta$ increases. However, a larger value of $\beta$ requires more time to converge to the stationary state. For example, the curve of $\beta=0.01$ in Figure \ref{fig: beta}(a) oscillates around $2160$ within only one second, while the curve of $\beta=0.1$ goes down continuously after even $50$ seconds. Let's consider an extreme case where $\beta$ is close to infinity. In this case, MH-SCCA may sink into some local optimum. It is almost impossible to reach the global optimum and achieve the stationary distribution. Therefore, in practical NFV systems, an appropriate value of $\beta$ needs to be selected to obtain both good time-averaged system performance and fast convergence.

\begin{figure}[!t]
	\centering
    \subfloat[Number of players.]{
    \label{subfig: num_p}
    \includegraphics[scale=.15]{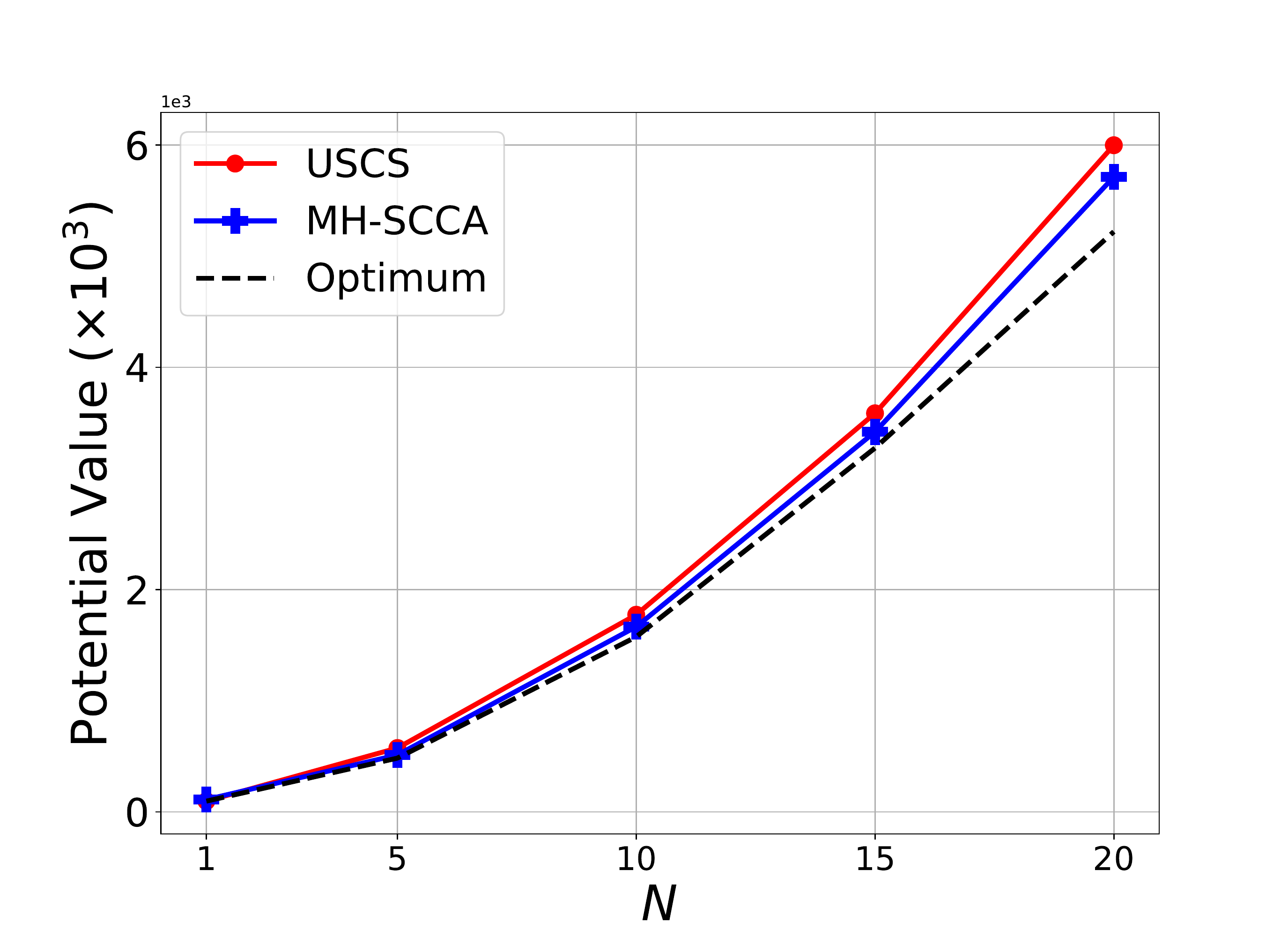}}
    \hfil
    \subfloat[Number of VNF types.]{
    \label{subfig: len_sc}
    \includegraphics[scale=.15]{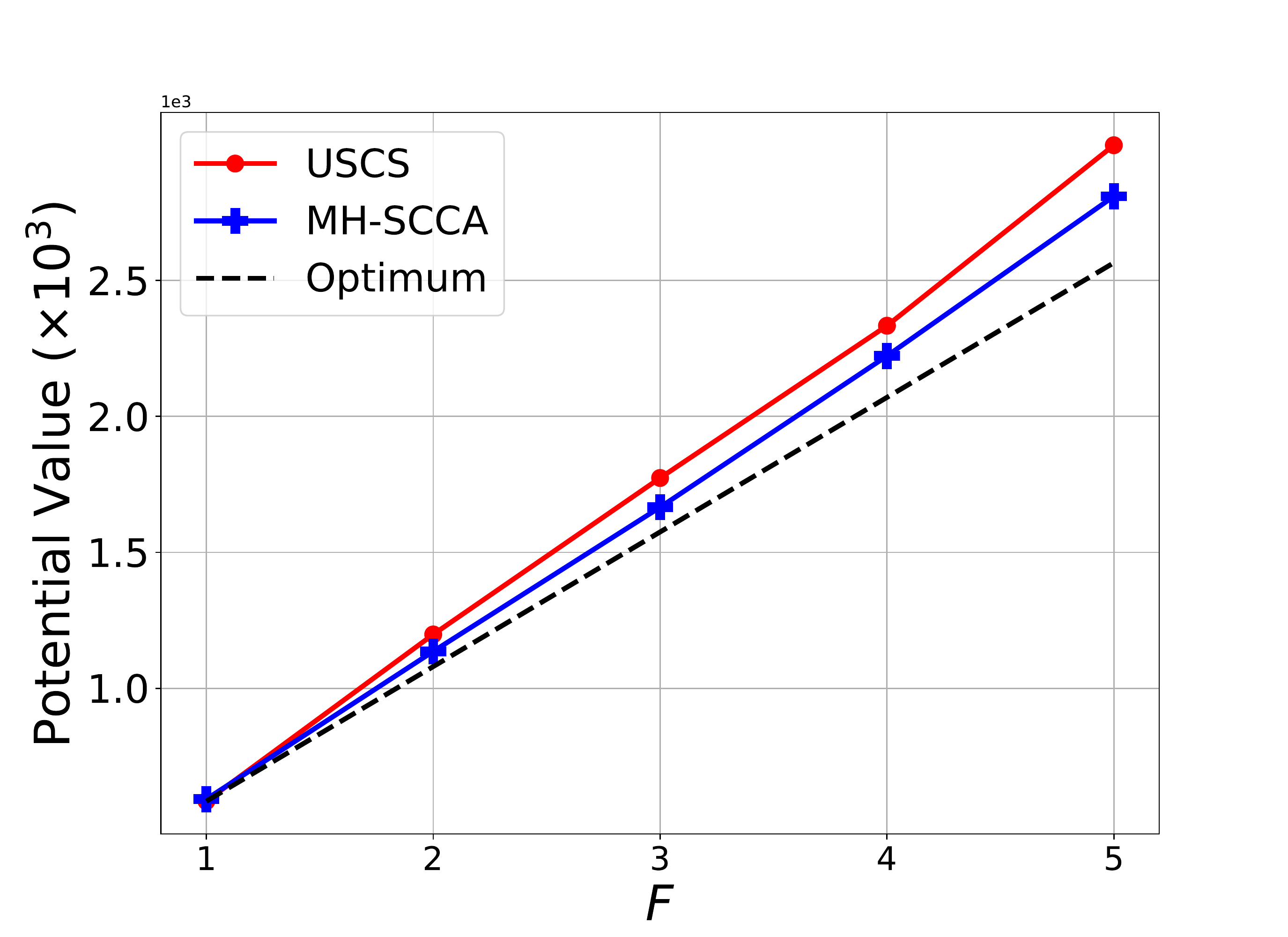}}
    \caption{Potential value of MH-SCCA under different settings of (a) number of players $N$ and (b) number of VNF types $F$.}
    \label{fig: N_F} 
\end{figure}

Figure \ref{fig: N_F} compares USCS and MH-SCCA under different values of (a) number of players $N$ and (b) number of VNF types $F$.\footnote{We assume that all players have the same service chain that consists of all types of VNFs in a specified order. Therefore, the number of VNF types is equivalent to the length of players' service chains.} We have the following observations: 
1) Given that there is only one player ($N=1$ in Figure \ref{fig: N_F}(a)) or the service chain consists of only one VNF ($F=1$ in Figure \ref{fig: N_F}(b)), USCS and MH-SCCA behave similarly because there is no local optima but only one global optimum. 
2) Under all other settings, MH-SCCA outperforms USCS. Moreover, the gap increases as the number of players $N$ (the number of VNF types $F$) increases. 

\begin{figure}[!t]
    \centering
    \subfloat[Player dynamics.]{
    \label{subfig: player dynamics}
    \includegraphics[scale=.15]{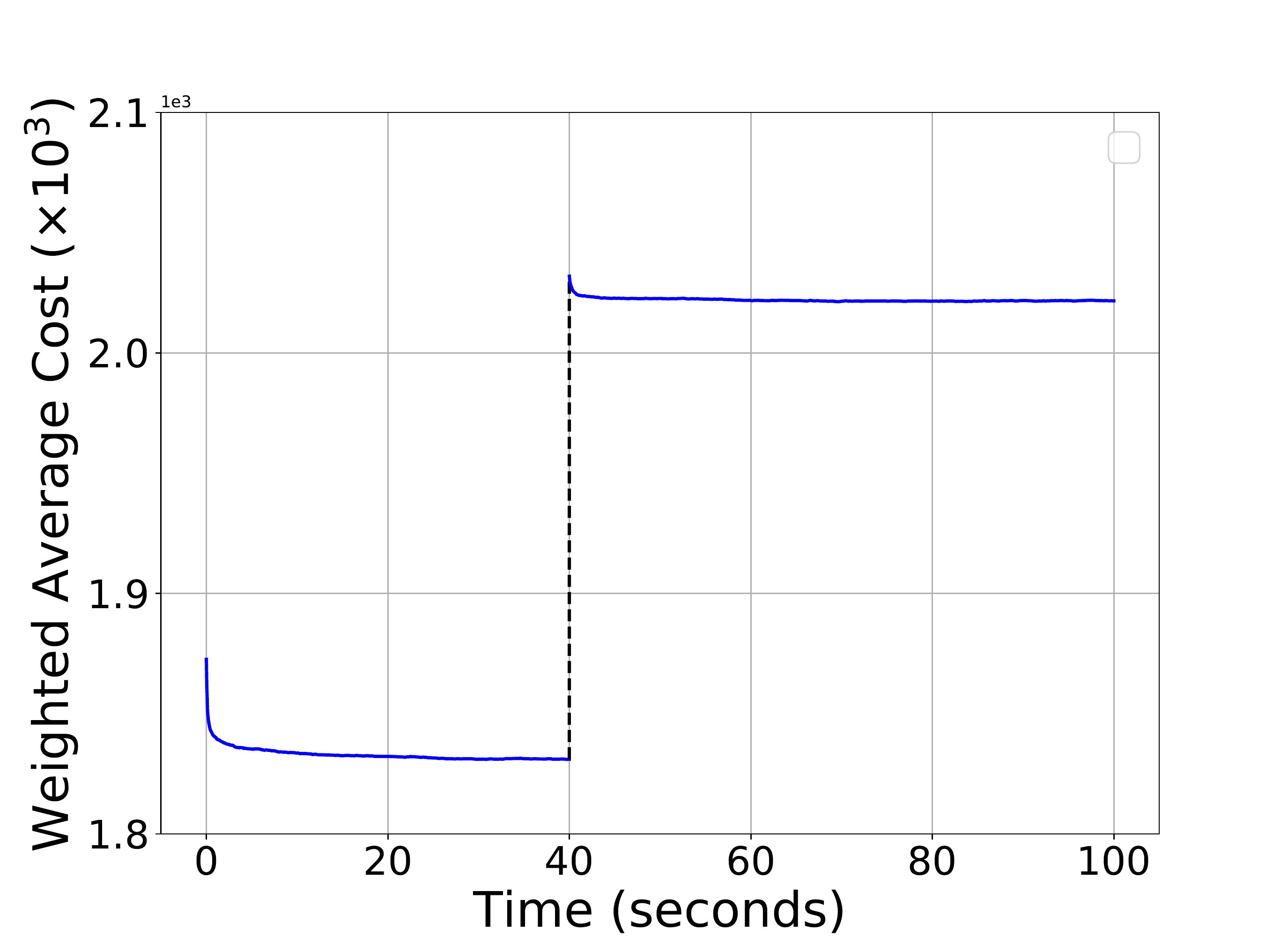}}
    \hfil
    \subfloat[VM dynamics.]{
    \label{subfig: VM dynamics}
    \includegraphics[scale=.15]{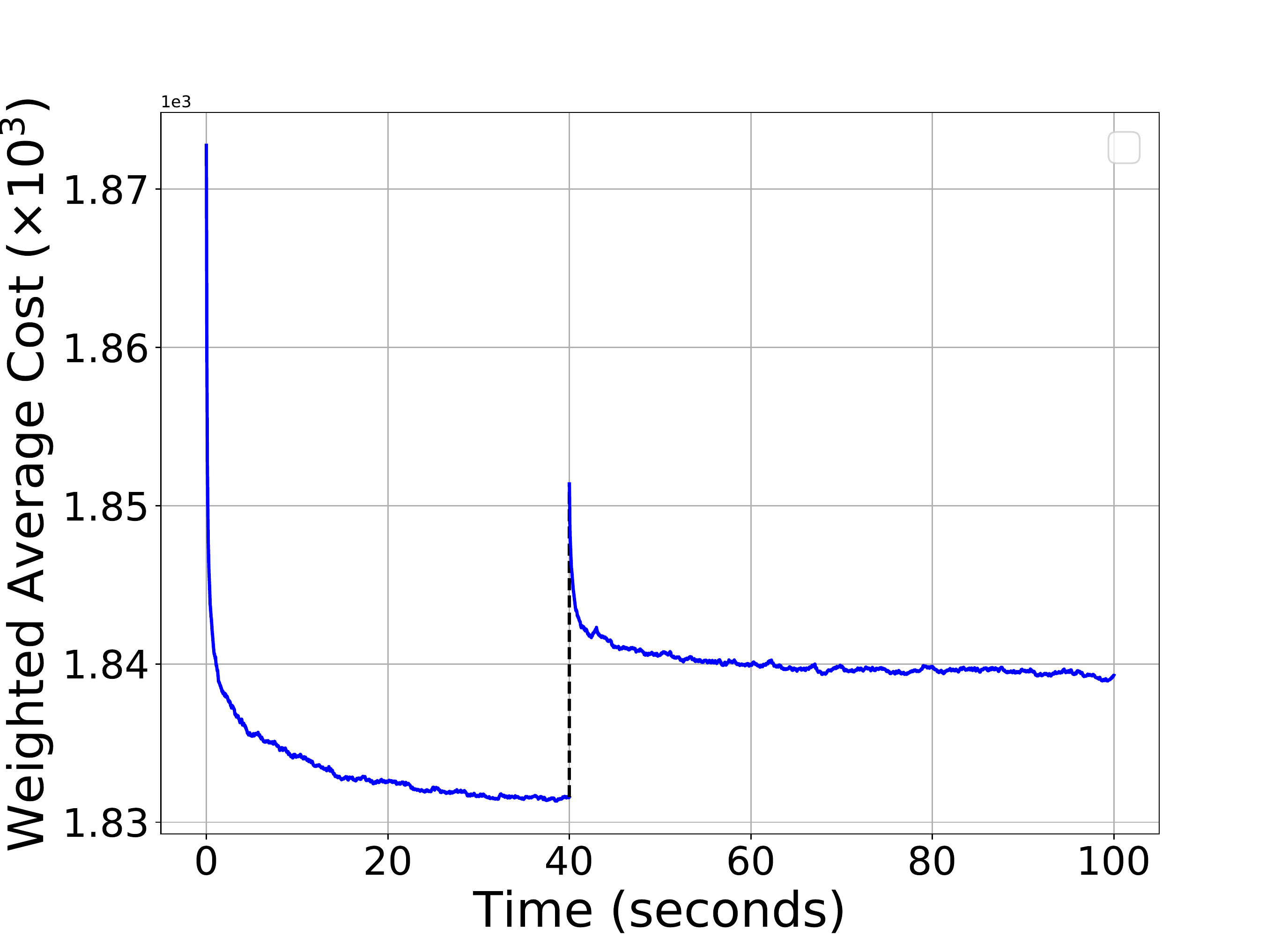}}
    \caption{Weighted average cost v.s. time when (a) a new player joins and (b) a VM fails at the $40$-th second.}
    \label{fig: dynamics} 
\end{figure}

\subsection{Adaptivity}
In practice, users often join and leave the NFV system dynamically  \cite{shah2010dynamics}. Moreover, a VM may break down unexpectedly due to various reasons such as network issues and power outages  \cite{birke2014failure}. 
Therefore, we investigate the adaptivity of MH-SCCA in response to such dynamics. 
    
We run the simulation for $100$ seconds, and a new player with traffic rate $10$Mbps will join at the $40$-th second. 
Once the new player joins, it picks a strategy from its strategy space randomly and starts an exponential count-down clock of rate $\mu$. 
Figure \ref{fig: dynamics}(a) presents the corresponding performance of MH-SCCA. We see that when the new player joins, the weighted average cost increases immediately and sharply. This is because the arrival of the new player increases the congestion costs of some players. Then, within a short period (less than five seconds), the weighted average cost decreases and converges again. Results from Figure \ref{fig: dynamics}(a) show the rapid adaptivity of MH-SCCA when faced with player dynamics. 
    
We also consider possible VM failures in the simulation. Specifically, We run the simulation for $100$ seconds, and turn off a specific VM at the $40$-th second. If one VM breaks down, all traffics on it will be migrated to another randomly selected VM of the same VNF type. Figure \ref{fig: dynamics}(b) shows that when the VM is down, the weighted average cost rises up immediately because the random migration may cause imbalance and incur high congestion cost. 
Then, within a short period (less than twenty seconds), the weighted average cost converges again. Results from Figure \ref{fig: dynamics}(b) show the rapid adaptivity of MH-SCCA when faced with VM dynamics. 

\subsection{Failures}
In this part we investigate how different failure probabilities affect the performance of MH-SCCA. 
Table \ref{tab: failures} shows the weighted average cost under different values of $\bar{\gamma}_{\text{(VM)}}$ and $\bar{\gamma}_{\text{(user)}}$ while the number of VNF types (service chain length) is $F = 2$. From this table, we see that the weighted average cost increases from left to right, and from top to bottom. In summary, a larger failure probability (whether player or VM) incurs a higher weighted average cost. 

\begin{table}
	\scriptsize
    \centering
    \caption{Weighted average cost under different values of $\bar{\gamma}_{\text{(VM)}}$ and $\bar{\gamma}_{\text{{\upshape (user)}}}$ with $F=2$.}
    \begin{tabular}{|c|c|c|c|c|c|c|}
         \hline
         \diagbox{$\bar{\gamma}_{\text{(user)}}$}{$\bar{\gamma}_{\text{(VM)}}$} & $0$ & $0.1$ & $0.2$ & $0.3$ & $0.4$ & $0.5$  \\
         \hline
         $0$ & $117.3$ & $1045$ & $1875$ & $2607.5$ & $3242.2$ & $3779.3$ \\
         \hline
         $0.1$ & $602.3$ & $1437.9$ & $2185.5$ & $2845.2$ & $3416.8$ & $3900.6$ \\
         \hline
         $0.2$ & $1088.2$ & $1831.5$ & $2496.5$ & $3083.2$ & $3591.8$ & $4022.1$ \\
         \hline
         $0.3$ & $1575$ & $2225.8$ & $2808$ & $3321.8$ & $3767$ & $4143.8$ \\
         \hline
         $0.4$ & $2062.5$ & $2620.6$ & $3120$ & $3560.6$ & $3942.5$ & $4265.6$ \\
         \hline
         $0.5$ & $2550.7$ & $3016.1$ & $3432.5$ & $3799.9$ & $4118.3$ & $4387.7$ \\ 
         \hline
    \end{tabular}
    \label{tab: failures}
\end{table}

\begin{figure}[!t]
    \centering
    \subfloat[$F = 2$.]{
    \label{subfig: failure_2} 
    \includegraphics[scale=.14]{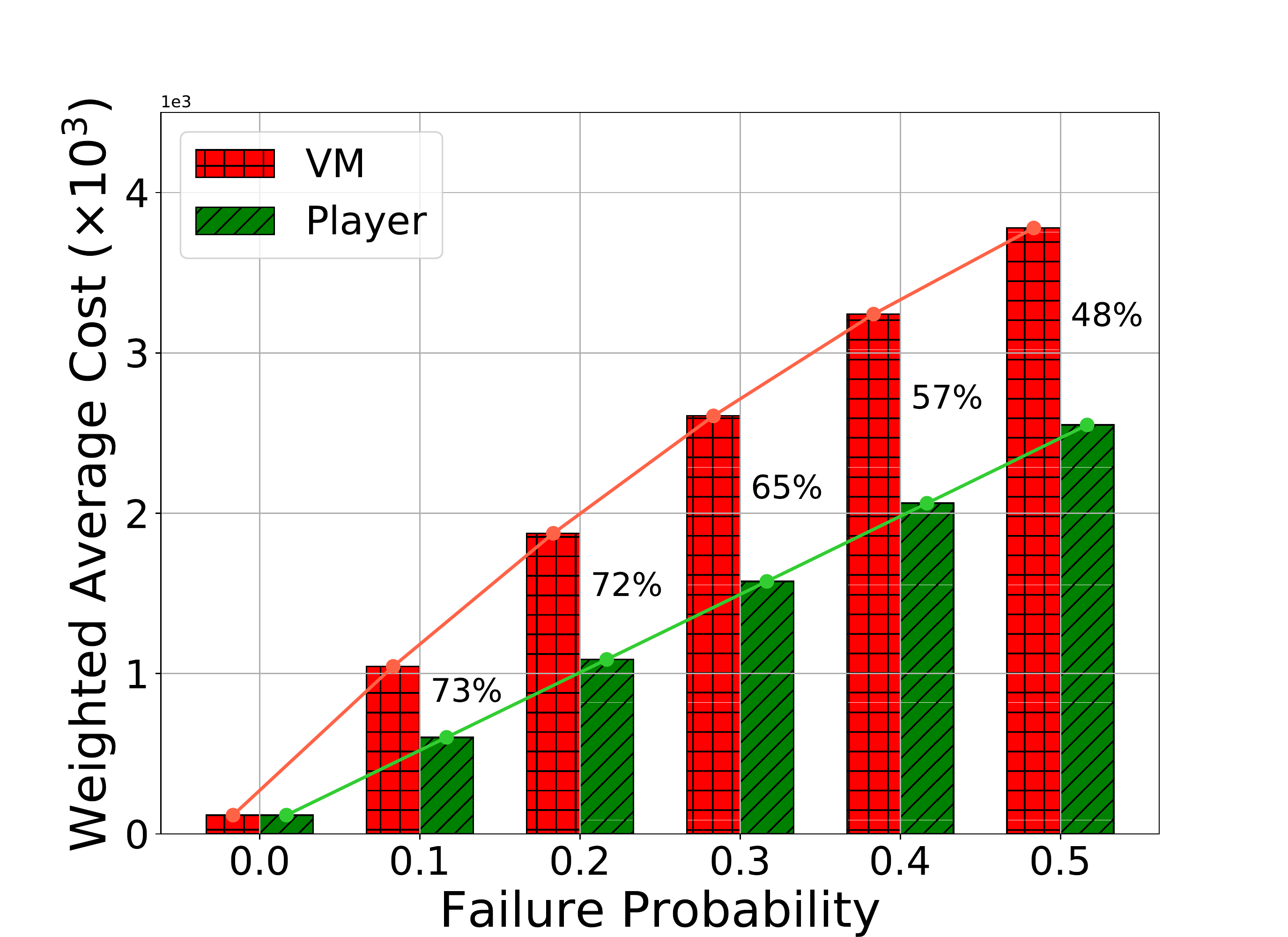}}
    \subfloat[$F = 3$.]{
    \label{subfig: failure_3}
    \includegraphics[scale=.14]{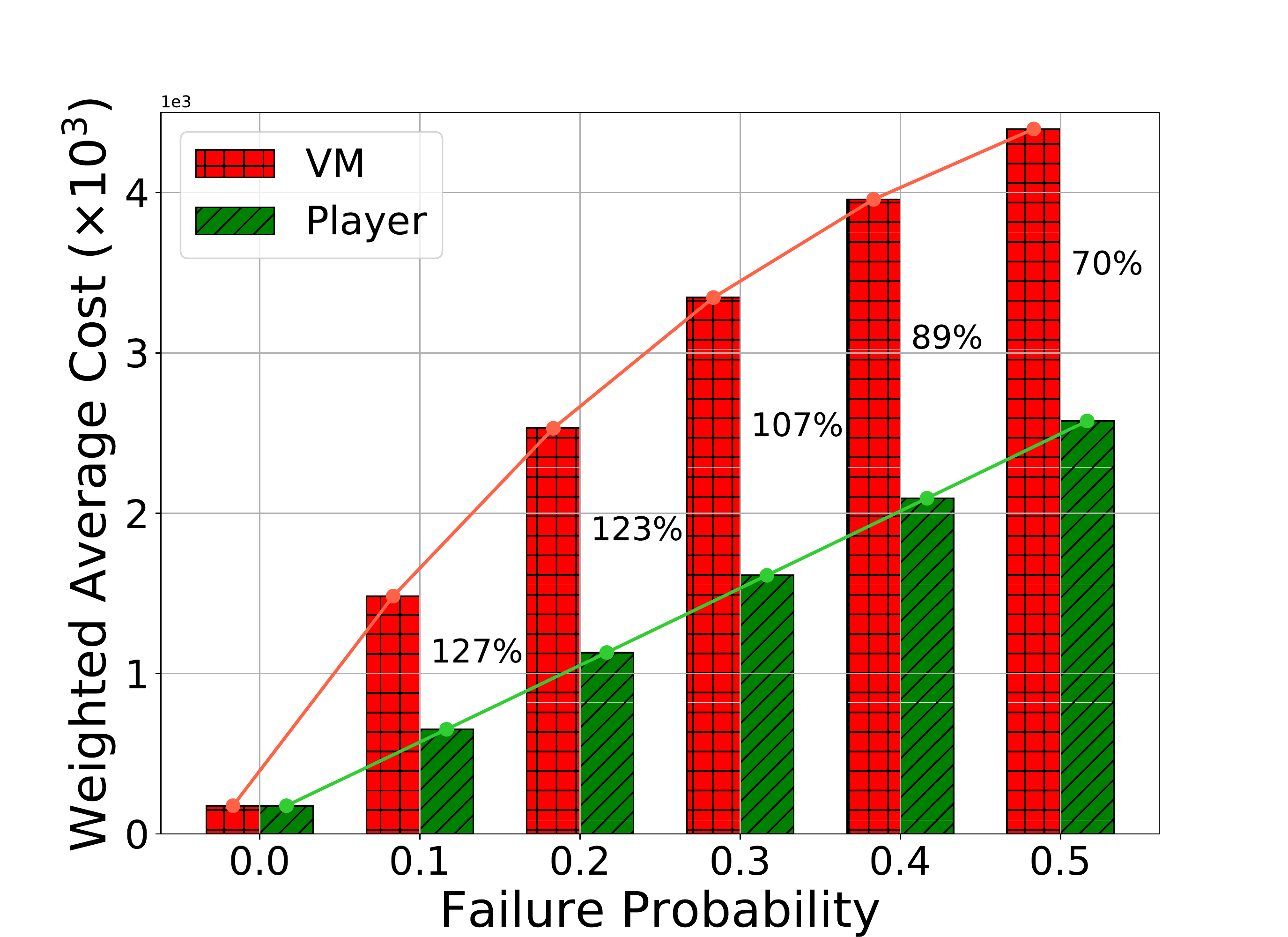}}
    \caption{Weighted average cost v.s. failure probability of player (green stripped bar) and VM under different values of $F$.}
    \label{fig: failures} 
\end{figure}

In Figure \ref{fig: failures}(a), we present the first row and the first column of Table \ref{tab: failures}. The red meshed bars represent the case when $\bar{\gamma}_{\text{(VM)}}$ increases from $0$ to $0.5$ and $\bar{\gamma}_{\text{(user)}}$ is set as $0$ by default (the first row). Similarly the green stripped bars represent the case when $\bar{\gamma}_{\text{(user)}}$ ranges from $0$ to $0.5$ and $\bar{\gamma}_{\text{(VM)}} = 0$ (the first column). 
Figure \ref{fig: failures}(b) presents similar results but under the setting of  $F=3$. 

From Figure \ref{fig: failures}, we see that the red meshed bars are always higher than the corresponding green stripped bars. 
In other words, with the same failure probability, the impact of VM failures on the weighted average cost is greater than that of player failures. 
The reason is that multiple VMs are employed to serve a single player, magnifying the impact of VM failures. Comparing Figure \ref{fig: failures}(a) and \ref{fig: failures}(b), we find that with a fixed failure probability (\textit{e.g.}, $0.1$), the gap between the red and green bars in Figure \ref{fig: failures}(b) ($127\%$) is larger than that in Figure \ref{fig: failures}(a) ($73\%$). The reason is that the impact of VM failures aggravates further as the length of service chains grows. 

\section{Conclusion}\label{sec: conclusion}
In this paper, we studied the service chain composition problem in NFV systems with both user and virtual machine failures. 
By formulating the problem from the perspective of non-cooperative game, we showed that such a game is a weighted potential game. 
Then we proposed MA-SCCA, a distributed scheme that guides the chain composition of different users in the system to the Nash equilibrium state with near-optimal request latencies and network congestion.
Moreover, by adopting Metropolis-Hastings methods, we proposed MH-SCCA, a variant of MA-SCCA, to further accelerate the composition process. In addition, we devised two novel learning-aided schemes, namely DRL-SCCA and MCTS-SCCA, based on DRL and MCTS techniques, respectively.
We conducted both theoretical analysis and numerical simulations to demonstrate the effectiveness of our proposed schemes.
The simulation results also validate the adaptivity of MH-SCCA when faced with network dynamics. 


\bibliographystyle{IEEEtran}
\bibliography{IEEEabrv, references}

\section*{Appendices}
\subsection{Proof of Theorem \ref{theorem: potential}}
We consider two states $\mathbf{w} = (\mathbf{w}_i, \mathbf{w}_{-i})$ and $\mathbf{w}' = (\mathbf{w}'_i, \mathbf{w}_{-i})$ that involves only player $i$'s strategy change. By the definition of cost function in (\ref{eq: cost_function}), we have
\begin{equation} \label{eq: proof1_step1}
\begin{split}
    &{c}_i(\mathbf{w}) - {c}_i(\mathbf{w}')\\
    =&~ \gamma_{\text{(user)}} \cdot \gamma_{\text{(VM)}}^{F_i} \cdot \left(\alpha c_i^{(\text{L})}(\mathbf{w}_i) - \alpha c_i^{(\text{L})}(\mathbf{w}'_i) \right) \\
    &+ \gamma_{\text{(user)}} \cdot \gamma_{\text{(VM)}}^{F_i} \cdot \left( {c}_i^{(\text{C})}(\mathbf{w}) - {c}_i^{(\text{C})}(\mathbf{w}') \right).
\end{split}
\end{equation}
By the definition of $\Phi$ in (\ref{eq: potential}), we have
\begin{equation}\label{eq: proof1_step2}
\begin{split}
    &\Phi(\mathbf{w}) - \Phi(\mathbf{w}') = 2 \gamma_{\text{(user)}} \cdot \lambda_i \left[ \alpha c_i^{(\text{L})}(\mathbf{w}_i) - \alpha c_i^{(\text{L})}(\mathbf{w}'_i) \right] \\
    & + \sum_{v \in \mathcal{V}} \Bigg[ \bigg(\sum_{k \in \mathcal{N}_v(\mathbf{w})} \gamma_{\text{(user)}} \cdot \lambda_k \bigg)^2 - \bigg(\sum_{k \in \mathcal{N}_v(\mathbf{w'})} \gamma_{\text{(user)}} \cdot \lambda_k \bigg)^2 \Bigg].
\end{split}
\end{equation}
Next, we focus on the second summation term on the right-hand side in (\ref{eq: proof1_step2}). The set of VMs $\mathcal{V}$ can be divided into the following subsets. 
\begin{itemize}
    \item[$\diamond$] $\mathcal{V}_1$: VMs adopted by player $i$ in $\mathbf{w}_i$ but not in $\mathbf{w}'_i$; 
    \item[$\diamond$] $\mathcal{V}_2$: VMs adopted in $\mathbf{w}'_i$ but not in $\mathbf{w}_i$; 
    \item[$\diamond$] $\mathcal{V}_3$: VMs that are both adopted in $\mathbf{w}_i$ and $\mathbf{w}'_i$;
    \item[$\diamond$] $\mathcal{V}_4$: VMs that are adopted neither in $\mathbf{w}_i$ nor $\mathbf{w}'_i$. 
\end{itemize}
Note that we always have $|\mathcal{V}_1| = |\mathcal{V}_2|$ because for any VM $v$ such that $v \in \mathbf{w}_i$ and $v \not \in \mathbf{w}'_i$, there must exist another VM $v'$ such that $v' \in \mathbf{w}'_i,$ and $v' \not \in \mathbf{w}_i$. Therefore, we have
\begin{equation}\label{eq: proof1_step3}
\begin{split}
    &\sum_{v \in \mathcal{V}} \Bigg[ \bigg(\sum_{k \in \mathcal{N}_v(\mathbf{w})} \gamma_{\text{(user)}} \lambda_k \bigg)^2 - \bigg(\sum_{k \in \mathcal{N}_v(\mathbf{w'})} \gamma_{\text{(user)}} \lambda_k \bigg)^2 \Bigg] \\
    =&~ \sum_{v \in \mathcal{V}_1} \Bigg[ \bigg(\sum_{k \in \mathcal{N}_v(\mathbf{w})} \gamma_{\text{(user)}} \lambda_k \bigg)^2 - \bigg(\sum_{k \in \mathcal{N}_v(\mathbf{w})} \gamma_{\text{(user)}} \lambda_k -  \gamma_{\text{(user)}}\lambda_i \bigg)^2 \Bigg]\\
    &+ \sum_{v \in \mathcal{V}_2} \Bigg[ \bigg(\sum_{k \in \mathcal{N}_v(\mathbf{w}')} \gamma_{\text{(user)}} \lambda_k - \gamma_{\text{(user)}} \lambda_i \bigg)^2 - \bigg(\sum_{k \in \mathcal{N}_v(\mathbf{w}')} \gamma_{\text{(user)}} \lambda_k \bigg)^2 \Bigg]\\
    =&~ 2\gamma_{\text{(user)}}\lambda_i \Bigg[\sum_{v \in \mathcal{V}_1} \bigg(\sum_{k \in \mathcal{N}_v(\mathbf{w})} \gamma_{\text{(user)}} \lambda_k \bigg) -  \sum_{v \in \mathcal{V}_2} \bigg(\sum_{k \in \mathcal{N}_v(\mathbf{w}')} \gamma_{\text{(user)}} \lambda_k \bigg)  \Bigg]\\
    =&~ 2\gamma_{\text{(user)}}\lambda_i \Bigg[\sum_{v \in \mathcal{R}_1} \sum_{k \in \mathcal{N}_v(\mathbf{w})/\{i\}} \gamma_{\text{(user)}} \lambda_k  -  \sum_{v\in \mathcal{V}_2} \sum_{k \in \mathcal{N}_v(\mathbf{w}')/\{i\}} \gamma_{\text{(user)}} \lambda_k  \Bigg]\\
    =&~ 2\gamma_{\text{(user)}}\lambda_i \Bigg[\sum_{v \in \mathcal{V}_1} \delta_{i, v}(\mathbf{w})  -  \sum_{v \in \mathcal{V}_2} \delta_{i, v}(\mathbf{w}) \Bigg]\\
    =&~ 2\gamma_{\text{(user)}}\lambda_i \left({c}_i^{(\text{C})}(\mathbf{w}) - {c}_i^{(\text{C})}(\mathbf{w}') \right).
\end{split}
\end{equation}
        
Combining (\ref{eq: proof1_step1}), (\ref{eq: proof1_step2}) and (\ref{eq: proof1_step3}), we obtain
\begin{equation}\label{proof1_step4}
\begin{split}
    \Phi(\mathbf{w}) - \Phi(\mathbf{w}') = 2\lambda_i \gamma_{\text{(VM)}}^{-F_i} \left({c}_i(\mathbf{w}) - {c}_i(\mathbf{w}') \right).
\end{split}
\end{equation}
\IEEEQED

\subsection{Proof of Theorem \ref{theorem: gap}}
Recall that $\mathbf{w}^*$ denotes the solution to problem (\ref{eq: problem_initial}), \textit{i.e.}, $\mathbf{w}^* = {\arg \min}_{\mathbf{w} \in \mathcal{W}} ~\Phi(\mathbf{w})$. Therefore, we have
\begin{equation}\label{eq: gap_proof}
    \exp \left[-\beta \Phi(\mathbf{w}^*)\right] \leq \sum_{\mathbf{w} \in \mathcal{W}} \exp [-\beta \Phi(\mathbf{w})] \leq |\mathcal{W}| \exp \left[-\beta \Phi(\mathbf{w}^*)\right]. 
\end{equation}
Next, we define
\begin{equation}\label{eq: g}
    g_{\beta} = -\frac{1}{\beta} \log \Big(\sum_{\mathbf{w} \in \mathcal{W}} \exp [-\beta \Phi(\mathbf{w})] \Big).
\end{equation}
By combining (\ref{eq: gap_proof}) and (\ref{eq: g}), we have
\begin{equation}
\begin{split}
     \Phi(\mathbf{w}^*) - \frac{1}{\beta} \log|\mathcal{W}| \leq g_{\beta} \leq \Phi(\mathbf{w}^*). 
\end{split}
\end{equation}
According to the property of conjugate function  \cite{boyd2004convex}, we know that $g_{\beta}$ equals the optimal value of problem (\ref{eq: problem_approx}). Therefore, the optimality gap is bounded by
\begin{equation}
    \frac{1}{\beta} \log|\mathcal{W}| = \frac{1}{\beta} \log \Big(\prod_{i\in \mathcal{N}} M^{F_i} \Big) 
    \leq\ \frac{1}{\beta} FN\log M.
\end{equation}
\IEEEQED

\subsection{Proof of Theorem \ref{theorem: chain}}
To show the time-reversibility of the Markov chain, we first show that it satisifies the following properties.
\begin{itemize} 
    \item[$\diamond$] \textit{The topology of the Markov chain is connected.} In the topology, the Hamming distance $H_{\mathbf{w}, \mathbf{w}'}$ equals the number of transitions needed to reach state $\mathbf{w}'$ from $\mathbf{w}$. From the definition of Hamming distance in (\ref{eq: hamming}), we conclude that $H_{\mathbf{w}, \mathbf{w}'} \leq N$ for any two states $\mathbf{w}$, $\mathbf{w}'$.
    Therefore, any state pairs are reachable within $N$ transitions, thus the designed topology is connected.
    \item[$\diamond$] \textit{Given any state $\mathbf{w} \in \mathcal{W}$, the probabilities of transitions out of it sum to one.} Specifically,
    \begin{equation}
    \begin{split}
        &\sum_{\mathbf{w}' \in \mathcal{W}} p_{\mathbf{w}, \mathbf{w}'} = \sum_{\substack{\mathbf{w}' \in \mathcal{W} \\ H_{\mathbf{w}, \mathbf{w}'}=0}} p_{\mathbf{w}, \mathbf{w}'} + \sum_{\substack{\mathbf{w}' \in \mathcal{W} \\ H_{\mathbf{w}, \mathbf{w}'}=1}} p_{\mathbf{w}, \mathbf{w}'}\\
        =& ~\frac{1}{N} \sum_{i\in \mathcal{N}} \frac{\exp [-\beta \Phi(\mathbf{w})]}{\sum_{\bar{\mathbf{w}} \in \mathcal{A}_i} \exp [-\beta \Phi(\bar{\mathbf{w}})]}\\
        &+ \sum_{i\in \mathcal{N}} \sum_{\substack{\hat{\mathbf{w}} \in \mathcal{A}_i\\ \hat{\mathbf{w}} \neq \mathbf{w}}} \frac{1}{N} \frac{\exp [-\beta \Phi(\hat{\mathbf{w}})]}{\sum_{\bar{\mathbf{w}} \in \mathcal{A}_i} \exp [-\beta \Phi(\bar{\mathbf{w}})] }\\
        =& ~\frac{1}{N} \sum_{i\in \mathcal{N}} \frac{\sum_{\hat{\mathbf{w}} \in \mathcal{A}_i} \exp [-\beta \Phi(\hat{\mathbf{w}})]}{\sum_{\bar{\mathbf{w}} \in \mathcal{A}_i} \exp [-\beta \Phi(\bar{\mathbf{w}})]} = 1.
    \end{split}
    \end{equation}

    \item[$\diamond$] \textit{The transition probabilities between any two states satisfy the detailed balance equation.} First, the detailed balance equation is naturally satisfied for state self transition. Next, for any two connected states $\mathbf{w}$ and $\mathbf{w}'$, by (\ref{eq: opt_prob}) and (\ref{eq: trans_prob_chain}), we have
    \begin{equation}
    \begin{split}
        &\pi^*_{\mathbf{w}} p_{\mathbf{w}, \mathbf{w}'} \\
        =& \frac{\exp [-\beta \Phi(\mathbf{w})]}{\sum_{\bar{\mathbf{w}} \in \mathcal{W}} \exp [-\beta \Phi(\bar{\mathbf{w}})]} \cdot \frac{1}{N} \frac{\exp [-\beta \Phi(\mathbf{w}')]}{\sum_{\bar{\mathbf{w}} \in \mathcal{A}_i} \exp [-\beta \Phi(\bar{\mathbf{w}})] }\\
        =& \frac{\exp [-\beta \Phi(\mathbf{w}')]}{\sum_{\bar{\mathbf{w}} \in \mathcal{W}} \exp [-\beta \Phi(\bar{\mathbf{w}})]} \cdot \frac{1}{N} \frac{\exp [-\beta \Phi(\mathbf{w})]}{\sum_{\bar{\mathbf{w}} \in \mathcal{A}_i} \exp [-\beta \Phi(\bar{\mathbf{w}})] }\\
        =& \pi^*_{\mathbf{w}'} p_{\mathbf{w}', \mathbf{w}}.
    \end{split}
    \end{equation}
\end{itemize}
Therefore, the designed Markov chain is irreducible. Moreover, it is also aperiodic because of the existence of self-transitions. 
Besides, it also satisfies the detailed balance equations. 
Hence, according to Chapter 11 in  \cite{blitzstein2014introduction}, we say the resultant Markov chain is time-reversible with its stationary distribution as (\ref{eq: opt_prob}). \\
\IEEEQED
    
\subsection{Proof of Theorem \ref{theorem: performance}}
We first focus on bounding the mixing time of MH-SCCA. 
In MH-SCCA, we denote the Markov chain's transition probability matrix by $P$. By (\ref{eq: trans_prob_chain}) and (\ref{eq: accept_prob}), for two states $\mathbf{w}$ and $\mathbf{w}'$ that differ at player $i$'s strategy, we have 
\begin{equation}
    P_{\mathbf{w}, \mathbf{w}'} = q_{\mathbf{w}, \mathbf{w}'} a_{\mathbf{w}, \mathbf{w}'} = \frac{1}{N|\mathcal{W}_i|} \min\left[\frac{\exp(-\beta \Phi(\mathbf{w}'))}{\exp(-\beta \Phi(\mathbf{w}))}, 1\right].
\end{equation} 

Then, we construct a new discrete-time Markov chain with transition probability matrix $P^{\text{lazy}} = \frac{1}{2} (I + P)$ where $I$ is the identity matrix. The self transition probability $P^{\text{lazy}}_{\mathbf{w}, \mathbf{w}} = \frac{1}{2} + P_{\mathbf{w}, \mathbf{w}} \geq \frac{1}{2}$, so we say such Markov chain is ``lazy". For any two different states $\mathbf{w}$ and $\mathbf{w}'$, the new transition probability is
$P^{\text{lazy}}_{\mathbf{w}, \mathbf{w}'} = \frac{1}{2} P_{\mathbf{w}, \mathbf{w}'}$. Such lazy Markov chain admits the same stationary distribution as the original one, because the equality $\pi^*_{\mathbf{w}} P_{\mathbf{w}, \mathbf{w}'} = \pi^*_{\mathbf{w}} P_{\mathbf{w}, \mathbf{w}'}$ implies $\pi^*_{\mathbf{w}} P^{\text{lazy}}_{\mathbf{w}, \mathbf{w}'} = \pi^*_{\mathbf{w}} P^{\text{lazy}}_{\mathbf{w}, \mathbf{w}'}$.  

We state that the mixing time of MH-SCCA is upper-bounded by the mixing time of the corresponding lazy Markov chain. Based on the Geometric method in  \cite{diaconis1991geometric}, we have
\begin{equation}\label{eq: bound_mix1}
    t_{\text{mix}}(\varepsilon) \leq
\frac{\log(\frac{1}{2\varepsilon})+\frac{1}{2}\log(\frac{1}{\pi_{\text{min}}}-1)}{\log(\frac{1}{\lambda_{2m}})},
\end{equation}
where $\pi_{\text{min}}$ is the minimum stationary probability, and $\lambda_{2m}$ is the second largest eigenvalue of the transition matrix $P^{\text{lazy}}$. To find an upper bound for the mixing time, we need to bound both $\pi_{\text{min}}$ and $\lambda_{2m}$. Firstly we give a lower bound of $\pi_{\text{min}}$:
\begin{equation}\label{eq: bound_pi}
\begin{split}
    \pi_{\text{min}} &= \min_{\mathbf{w} \in \mathcal{W}} \pi^*_{\mathbf{w}} = \frac{\min_{\mathbf{w} \in \mathcal{W}} \exp{[-\beta \Phi(\mathbf{w})]}}{\sum_{\bar{\mathbf{w}} \in \mathcal{W}} \exp{[-\beta \Phi(\bar{\mathbf{w}})]}} \\
    &\geq \frac{\exp(-\beta D)}{|\mathcal{W}|} \geq \frac{ \exp(-\beta D) }{M^{NF}},
\end{split}
\end{equation}
in which $D$  is the difference between the maximum and minimum potential values as defined in (\ref{def: D}).
Now we bound $\lambda_{2m}$ by Cheeger’s inequality  \cite{diaconis1991geometric}:
\begin{equation}\label{eq: buound_lambda1}
   \lambda_{2m} \leq 1-\frac{\varphi^2}{2},
\end{equation}
in which $\varphi$ is the ``conductance" of $P^{\text{lazy}}$ defined as
\begin{equation}
    \varphi \triangleq \min_{A \subset \mathcal{W}, \pi_{A} \in (0,\frac{1}{2}]} \frac{F(A,A^c)}{\pi_A},
\end{equation}
where $F(A,A^c) \triangleq \sum_{\mathbf{w} \in A, \mathbf{w'} \in A^c} \pi_{\mathbf{w}}^{*} P^{\text{lazy}}_{\mathbf{w},\mathbf{w'}}$ and 
$\pi_A \triangleq \sum_{\mathbf{w} \in A} \pi_{\mathbf{w}}^*$.

Next, we show a lower bound of $\varphi$. Specifically,
\begin{equation}\label{eq: bound_phi}
\begin{split}
    \varphi &\geq \min_{A \subset \mathcal{W}, \pi_{A} \in (0,\frac{1}{2}]} 2F(A,A^c) \\
    &\geq \min_{\mathbf{w} \neq \mathbf{w}', P^{\text{lazy}}_{\mathbf{w},\mathbf{w'}} > 0} 2\pi_{\mathbf{w}}^{*} P^{\text{lazy}}_{\mathbf{w},\mathbf{w'}} \\
    &\geq \min_{\mathbf{w} \neq \mathbf{w}', P_{\mathbf{w}, \mathbf{w}'} > 0} \pi_{\mathbf{w}}^{*} P_{\mathbf{w},\mathbf{w'}} \\
    &\geq \pi_{\text{min}} \frac{1}{N|\mathcal{W}_i|} \exp(-\beta D) \\
    &\geq \frac{ \exp(-2\beta D) }{NM^{F(N+1)}}. 
\end{split}
\end{equation}
Combining (\ref{eq: buound_lambda1}) and (\ref{eq: bound_phi}), we have
\begin{equation}\label{eq: bound_lambda2}
    \lambda_{2m} \leq 1 - \frac{ \exp(-4\beta D) }{2N^2 M^{2F(N+1)}}.
\end{equation}
Substituting (\ref{eq: bound_pi}) and (\ref{eq: bound_lambda2}) into (\ref{eq: bound_mix1}), we have that the mixing time of MH-SCCA is bounded by
\begin{equation}\label{eq: bound_mix2}
    t_{\text{mix}}(\varepsilon) \leq \frac{ \log(\frac{1}{2\varepsilon })+\frac{1}{2} FN \log{M}+\frac{1}{2} \beta D}{-\log[1-\frac{1}{2 N^2 M^{2F(N+1)}}\exp{(-4\beta D)}]}. 
\end{equation}
When it comes to MA-SCCA, we denote its transition probability matrix as $Q$. Then for two different states $\mathbf{w}$ and $\mathbf{w}'$,
\begin{equation}
    Q_{\mathbf{w}, \mathbf{w}'} = \frac{1}{N} \frac{\exp [-\beta \Phi(\mathbf{w}')]}{\sum_{\bar{\mathbf{w}} \in \mathcal{A}_i} \exp [-\beta \Phi(\bar{\mathbf{w}})]}
\end{equation}
and
\begin{equation}
    Q^{\text{lazy}}_{\mathbf{w}, \mathbf{w}'} = \frac{1}{2} Q_{\mathbf{w}, \mathbf{w}'}.
\end{equation}
Similar to (\ref{eq: bound_phi}), we derive the lower bound of the conductance of transition matrix $Q^{\text{lazy}}$ as 
\begin{equation}
\begin{split}
    \varphi &\geq \pi_{\text{min}} \frac{1}{N} \frac{\exp [-\beta \Phi(\mathbf{w}')]}{\sum_{\bar{\mathbf{w}} \in \mathcal{A}_i} \exp [-\beta \Phi(\bar{\mathbf{w}})] }\\
    &\geq \frac{ \exp(-2\beta D) }{NM^{F(N+1)}},
\end{split}
\end{equation}
which is the same as (\ref{eq: bound_phi}). 
Therefore, MA-SCCA has the same mixing time upper bound (\ref{eq: bound_mix2}) as MH-SCCA, thus the proof of Theorem \ref{theorem: performance} is completed. \\
\IEEEQED

\end{document}